\documentclass[useAMS,usenatbib]{mn2e}
\usepackage{natbib}
\bibpunct{(}{)}{;}{a}{,}{,}

\usepackage{graphicx}
\usepackage{helvet}
\usepackage{natbib}
\usepackage{setspace}
\usepackage{amssymb}
\usepackage{longtable}
\title[Variable stars]
{Variable stars in young open star cluster NGC 7380}

\author[Sneh Lata et al.]
       {Sneh Lata$^1$\thanks{E-mail: sneh@aries.res.in}, A. K. Pandey$^1$, Neelam Panwar$^2$, W. P. Chen$^{3}$ \\  
\newauthor  M. R. Samal$^{4}$, J. C. Pandey$^1$  \\
       $^1$Aryabhatta Research Institute of Observational Sciences, Manora Peak, Nainital 263002, Uttarakhand, India \\
       $^2$Department of Physics \& Astrophysics, University of Delhi-110007, India \\
       $^3$Institute of Astronomy, National Central University, 300 Jhongda Rd, Jhongli, Taoyuan Country 32054, Taiwan \\
       $^4$Aix-Marseille Universiti$\acute{e}$, CNRS, Laboratoire d'Astrophysique de Marseille UMR 7326, 13388, Marseille, France \\}

\date{Accepted ---------.
      Received ---------;
      }

\pagerange{\pageref{firstpage}--\pageref{lastpage}}

\def\LaTeX{L\kern-.36em\raise.3ex\hbox{a}\kern-.15em
    T\kern-.1667em\lower.7ex\hbox{E}\kern-.125emX}

\begin{document}

\label{firstpage}

\maketitle

\label{firstpage}
\begin{abstract}
We present 
time series photometry of 57 variable stars in the cluster
region NGC 7380. The association of these variable stars to the cluster NGC 7380 has been established 
on the basis of two colour diagrams and colour-magnitude diagrams.
Seventeen stars are found to be main-sequence variables, which are mainly 
B type stars and are classified as slowly pulsating B stars, $\beta$ Cep or 
$\delta$ Scuti stars. Some of them may belong to new class variables as discussed by Mowlavi et al. (2013) and Lata et al. (2014).
Present sample also contains 14 pre-main-sequence stars, whose ages and masses are found to be mostly $\lesssim$ 5 Myr and 
range 0.60 $\lesssim M/M_{\odot} \lesssim$ 2.30 and hence
 should be T-Tauri stars. About half of the weak line T-Tauri stars are found to be
fast rotators with a period of $\lesssim$ 2 days as compared to the classical T-Tauri stars.
Some of the variables belong to the field star population. 
\end{abstract}

\begin {keywords} 
Open  cluster:  NGC 7380  --
colour--magnitude diagram: Variables: pre-main sequence stars 
\end {keywords}

\section{Introduction}
NGC 7380 (RA=22h 47m 21s, Dec +58$^{\circ}$ 07$^\prime$ 54$^\prime$$^\prime$) is a young open cluster in the 
northern sky. 
Several studies of the NGC 7380 region based on photometric observations have already been carried out (e.g.,
Moffat 1971; Chen et al. 2011 and references therein). The members of the cluster were
identified by Baade (1983) on the basis of photometric observations by Moffat (1971).
They identified a few T-Tauri stars (TTSs) and Herbig Ae stars in the region. 
Mathew et al. (2010, 2012) found a number of pre-main-sequence (PMS) stars on the basis of optical
and near IR photometry. Chen et al. (2011) have carried out a detailed study of the region. Using the proper motion data they identified probable members of the cluster. They found that the reddening towards the direction of the cluster is variable with $E(B-V)_{min}$=0.5 mag and $E(B-V)_{max}$=0.7 mag and the cluster is located at a distance of 2.6$\pm$0.4 kpc. Using surface density map of 2MASS stars, Chen et al. (2011) also found that the cluster is elongated in the north-south direction with an average angular radius of 4 arcmin which corresponds to 3 pc at the distance of the cluster. They estimated the age of the cluster  as $\sim$ 4 Myr. The O-type binary star DH Cep is a member of the cluster which is in its late stage of clearing the surrounding material, and may have triggered the ongoing star formation in neighbouring molecular clouds which harbour young stars that are coeval and comoving but not gravitationally bound.  This cluster has been an interesting and important object because it contains a massive binary system DH Cep (=HD 215835; $V$ = 8.58) which is a double-lined, massive
spectroscopic binary having a pair of O5-6 V stars with an orbital
period of 2.111 days (Lines et al. 1986; Semeniuk 1991; Pearce 1949). This star is located at the center of the cluster and found to be an ionizing star for
the emission nebula (Underhill 1969).
Underhill (1969) found a very high
fraction of binaries in the region. Of the 10 stars observed by Underhill (1969), 6 are detected and 4 are suspected
spectroscopic binaries.

%*************************************************************************************************************************************************
\begin{figure*}
\includegraphics[width=17cm]{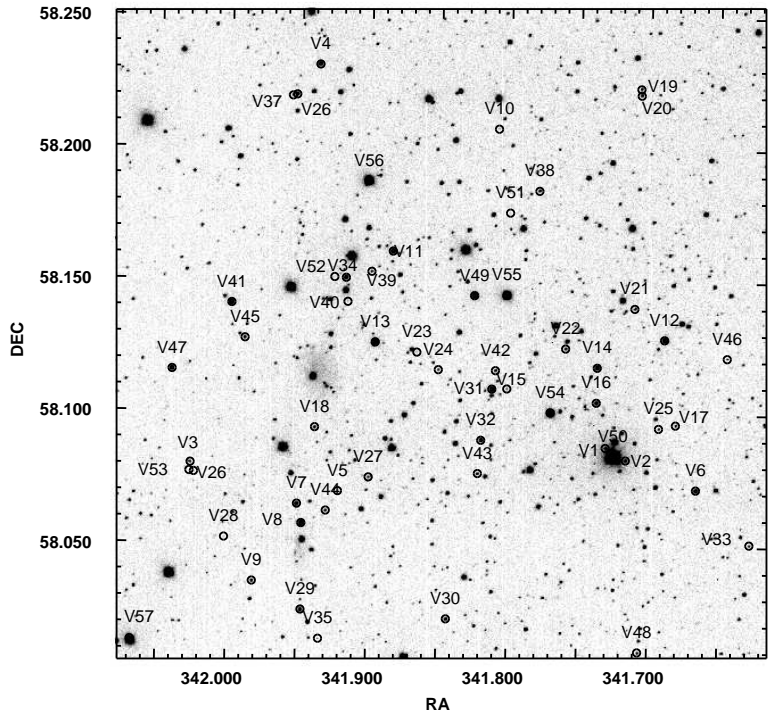}
\caption{ The observed region of NGC 7380 in V band. The variable candidates detected in the present work are encircled and labeled with numbers. The epoch of equatorial coordinates RA and DEC is J2000.0. }
\end{figure*}
%%*******************************************************************************************************************************
We find NGC 7380 an interesting object as it is an extremely young open cluster which contains several PMS stars, massive stars O/B type stars as well as other main-sequence (MS) stars.
Therefore we carried out photometric observations of the NGC 7380 region to search for variable stars.
Photometric variability in OB supergiants, early B-type stars, Be stars, mid to late B-type stars
occurs mostly due to the pulsations (Stankov \& Handler 2005; Kiriakidis et al. 1992; Moskalik \& Dziembowski 1992).
Pulsating variable stars expand and contract in a repeating cycle of size changes.
The different types of pulsating variables are distinguished by their periods of pulsation and the shapes of their light curves.
In PMS objects like TTSs (mass $\lesssim$ 3 $M_{\odot}$) photometric variations are believed to originate from several mechanisms like rotation of a star with an asymmetrical distribution of cool spots, variable hot spots or obscuration by circumstellar dust (see Herbst et al. 1994 and reference therein).
The Herbig Ae/Be stars (PMS stars having mass $\gtrsim$ 3 $M_{\odot}$) also show variability
as they move across the instability region in the Hertzsprung-Russell (HR) diagram on their way to the MS.
Several systematic studies of TTSs have been carried out which revealed different type of variabilities (Herbst et al. 1994). It is now well known that
some of PMS stars show periodic variability (e.g., Hillenbrand 2002; Schaefer 1983; Bouvier et al. 1993; Bouvier 1994; Percy et al. 2006, 2010).

 NGC 7380 has been monitored on 70 nights during October 2012 to February 2013      
to identify and characterize the variable stars in the region. In Section 2 we describe the observations, data reduction procedure, variable identification and period determination.  
In Section 3 we discuss association of the detected variables with the cluster using $(U-B)/(B-V)$ and $(J-H)/(H-K)$ two colour diagrams (TCDs) and the $V/(V-I)$ colour-magnitude diagram (CMD). Section 4 describes the age and mass
estimation of young stellar objects. 
In Section 5 we study the spectral energy distribution of identified
young stellar objects. 
Section 6 describes luminosity and temperature of the stars.
We characterize variable stars in Section 7. 
In section 8 we studied the effect of NIR excess on the rotation, whereas section 9 discusses the correlation among rotation, mass, age and amplitude.
We summarise our results in Section 10.

\section{Observations and Data Reduction}
The photometric observations of NGC 7380 were taken using the 0.81-m f/7 Ritchey-Chretien Tenagara automated telescope in southern Arizona. 
The telescope has a 1024$\times$1024 pixel SITe camera. Each pixel corresponds to 0.87 arcsec which yields a field of view of $\sim$ 14.8$\times$14.8 arcmin$^2$.
The observations were taken in $V$ and $I$ filters during 2012 October 19 to 2013 February 08 on 70 nights. 
Each night consists of 2 to 4 frames in $V$ and $I$ filters. On two nights 23 November 2012 and 24 November 2012 each had 
around 150 observations. The typical seeing (estimated from full width at half maximum of the point like stars) of the images was $\sim$2 to $\sim$3 arcsec. Bias and
twilight flats were also taken along with the target field. The observed region is shown in Fig. 1.

The preprocessing of the CCD images was performed using the IRAF\footnote{IRAF is distributed by the National Optical Astronomy Observatory, which is operated by the Association of Universities for Research in Astronomy (AURA) under cooperative agreement with the National Science Foundation.} and
the instrumental magnitude of the stars were obtained using the DAOPHOT package (Stetson 1987).
Details can be found in our earlier papers (Lata et al. 2011, 2012).
We have used the DAOMATCH (Stetson 1992) routine of DAOPHOT  to find the translation, rotation and scaling solutions between different
photometry files, whereas DAOMASTER (Stetson 1992) matches the point sources.
To remove frame-to-frame flux variation
due to airmass and exposure time, we used DAOMASTER programme
to get the corrected magnitude.
This task makes the mean flux level of each frame equal to the
reference frame by an additive constant.
The first target frame taken on 23 November 2012 has been considered as the reference frame. Present observations have 641 and 631 frames in the $V$ and $I$ band respectively and each frame corresponds to one photometry file.
We have considered only those stars for further study which have at least 100 observations in the $V$ and $I$.
 From the output of DAOMASTER, we get corrected magnitudes listed in a $.cor$ file. This file is further used to search for variable stars in the next section.

The standardization of the stars in the cluster field was carried out  with the help of the data given by Chen et al. (2011).
The instrumental magnitudes were transformed to the standard Johnson $V$ and Cousins $I$ system. The equations used for photometric calibration are given below:

%**************************************************************
\begin{eqnarray}
V-v =  (-0.001\pm0.002)\times (V-I) + (0.902\pm0.007)   \nonumber\\
V-I = (0.965\pm0.011)\times (v-i) + (1.201\pm0.006)     \nonumber
\end{eqnarray}
%**************************************************************
Where $V$ and $I$ are standard magnitude of stars and $v$ and $i$ are
instrumental magnitudes of stars in $V$ and $I$ filters. 
The mean value of the photometric error of the data was estimated using the observations of each star. The mean error as a function of instrumental magnitude is
shown in Fig. 2.
The mean error is found to be $\sim$ 0.01 mag at
 magnitude $\le$ 13 mag and increases to $\sim$0.1 mag at $\sim$17 mag.

\subsection{Variable identification and period determination}
The differential magnitudes ($\Delta$m) of stars in the sense variable minus comparison star were plotted against  Julian date (JD).
The probable variables were identified visually by inspecting the light curves. 
The visual inspection yields 57 variable candidates, of which  
 50 were detected in $V$ band and 54  in
$I$ band including the massive binary DH Cep (HD 215835). 
The brightness of these variables vary from short to long time scale
periods. The samples of short and long period variables are shown in Figs. 3 and 4, 
respectively.

The identification number, coordinates and present $VI$ photometric data for these variable stars
are given in Table 1.
The CCD pixel coordinates of these identified variables were converted to celestial coordinates (RA and DEC) for J2000 with the help of the CCMAP and CCTRAN
tasks in IRAF.

 The variability of stars nos 51 to 57 could only be detected in $I$ band. 
Stars no 54, 55, 56 and 57 got saturated in $V$ band. Star no 51 to 53 are too faint to detect in $V$ band.  
Similarly, variability for stars 23, 28 and 35 could not be detected in I band. 
Table 1 also lists  $NIR$ data taken from the 2MASS catalogue (Cutri et al. 2003).
The variable candidates identified in the present work are marked in Fig. 1.

In order to determine the probable periods of the variable stars we used the Lomb-Scargle (LS) periodogram (Lomb 1976; Scargle 1982). This method works well with unevenly sampled data. 
We used the algorithm
available at the Starlink\footnote[3]{http://www.starlink.uk} software
database.
We further confirmed the periods using NASA exoplanet archive periodogram service. We visually inspected the phased light curves and opted the period showing the best light curve.
The most probable periods with amplitude are listed in Table 1.
The light curves of variable stars are folded with their estimated periods.
The phased light curves will be discussed in Section 5.

\begin{figure}
\includegraphics[width=8cm]{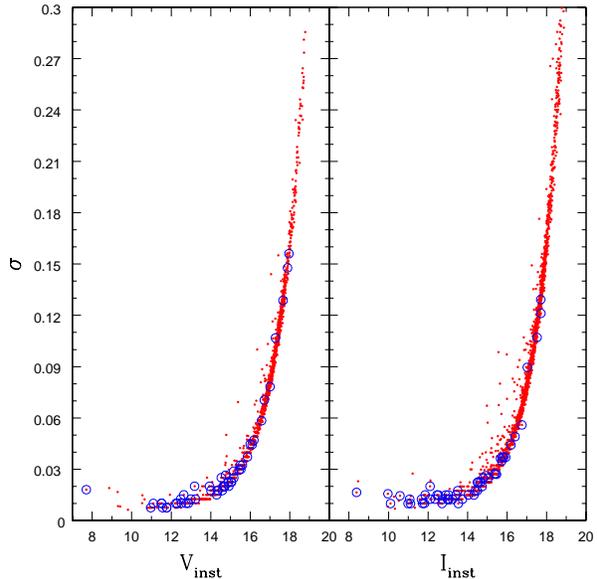}
\caption{Mean photometric errors given by DAOPHOT as a function of magnitude in $V$ and $I$.
Open circles represent
variable candidates.}
\end{figure}

\begin{figure*}
\includegraphics[width=18.cm, height=12cm]{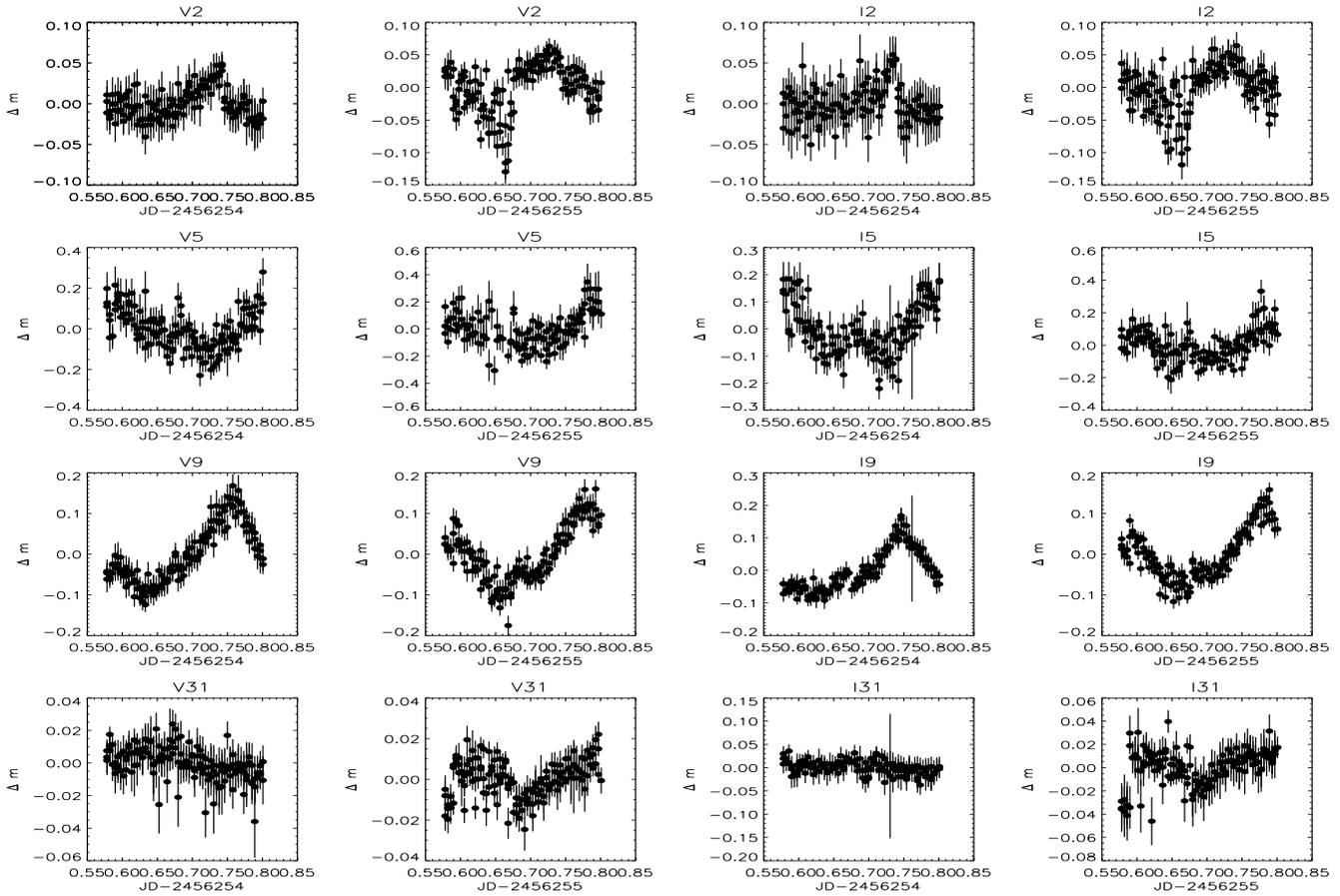}
\caption{The light curves of a few short period variable stars in $V$ and $I$ band identified in the
present work. The ID e.g., V2 and I2 represent variable no 2 in $V$ and $I$ bands, respectively. The $\Delta$m represents the differential magnitude. }
\end{figure*}
%*******************************************************************************************************************************
%%*******************************************************************************************************************************
\begin{figure*}
\includegraphics[width=18.cm, height=12cm]{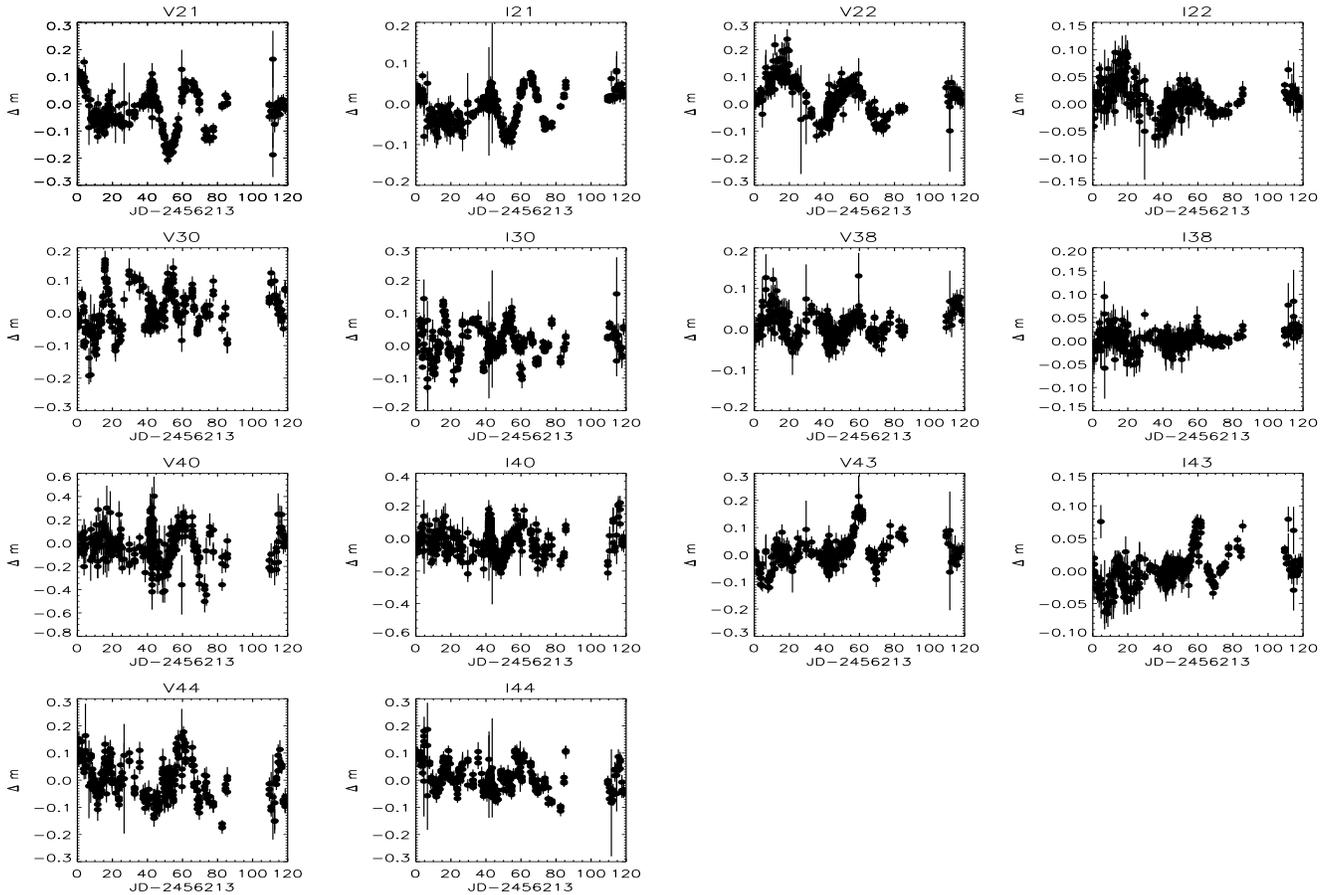}
\caption{The light curves of a few long period variable stars identified in the
present work, each with 70 nights observations. The $\Delta$m represents the differential magnitude. }
\end{figure*}
%*******************************************************************************************************************************

%*************************************************************************************************************************************************

%*************************************************************************************************************************************************
\begin{table*}
%\centering
\caption{The present photometric data, NIR data, period and amplitude of 57 variables in the region of NGC 7384. 
The NIR data have been taken from 2MASS point source catalogue (Cutri et al. 2003).
}
\tiny
\begin{tabular}{llllllllll}
\hline
ID&
$\alpha_{2000}$&
$\delta_{2000}$&
$V-I$&
$V$&
$J$&
$H$&
$K$&
[period]&
[Amp.]
\\
&
degree&
degree&
mag&
mag&
mag&
mag&
mag&
day&
mag

\\
\hline
 1  &  341.730639 & 58.087500  & 0.682$\pm$0.036 &16.079$\pm$0.029 & 14.949$\pm$0.041& 14.504$\pm$ 0.05 & 14.467$\pm$0.087 &  0.167&           0.108   \\      
 2  &  341.716194 & 58.082861  & 0.603$\pm$0.022 &14.847$\pm$0.020 & 13.725$\pm$0.029& 13.607$\pm$0.033 & 13.576$\pm$0.044 &  0.198&           0.027   \\      
 3  &  342.026556 & 58.080361  & 1.994$\pm$0.022 &16.442$\pm$0.032 & 12.550$\pm$0.031& 11.843$\pm$0.048 & 11.597$\pm$0.041 &  1.151&           0.100  \\       
 4  &  341.937500 & 58.231500  & 0.798$\pm$0.012 &13.226$\pm$0.010 & 11.770$\pm$0.022& 11.534$\pm$0.029 & 11.422$\pm$0.021 &  4.729&           0.146  \\       
 5  &  341.921167  &58.070111  & 1.391$\pm$0.049 &17.652$\pm$0.058 & 14.978$\pm$0.037& 14.480$\pm$0.054 & 14.382$\pm$0.08 &  0.325&           0.120   \\      
 6  &  341.665917 & 58.071778  & 1.033$\pm$0.010 &13.440$\pm$0.010 & 11.637$\pm$0.023& 11.330$\pm$ 0.03 & 11.258$\pm$0.023 &  1.255&           0.015  \\       
 7  &  341.950222 & 58.065167  & 0.959$\pm$0.017 &14.144$\pm$0.013 & 12.253$\pm$0.03 & 11.879$\pm$0.036 & 11.730$\pm$0.031 &  0.249&           0.015  \\       
 8  &  341.946917 & 58.057806  & 1.602$\pm$0.009 &12.428$\pm$0.010 &  9.580$\pm$0.023&  8.894$\pm$0.023 &  8.794$\pm$0.021 &  0.143&           0.015  \\       
 9  &  341.981556 & 58.035750  & 1.241$\pm$0.022 &15.372$\pm$0.025 & 13.355$\pm$0.027& 12.905$\pm$0.032 & 12.835$\pm$0.035 &  0.259&           0.076  \\       
 10 &  341.809000  &58.207806  &1.422$\pm$0.089  &18.126$\pm$0.106 & 15.628$\pm$0.072& 15.223$\pm$0.082 & 14.922$\pm$0.134  & 0.234 &           0.160 \\       
 11 &  341.883778  &58.161111  &0.563$\pm$0.015  &12.674$\pm$0.008 & 11.652$\pm$0.023& 11.567$\pm$0.028 & 11.480$\pm$0.021  & 1.074 &           0.012 \\       
 12 &  341.689111  &58.128528  &0.728$\pm$0.020  &12.464$\pm$0.008 & 11.105$\pm$0.032& 10.898$\pm$0.042 & 10.666$\pm$0.032  & 1.423 &           0.018  \\      
 13 &  341.895778  &58.126639  &0.531$\pm$0.012  &11.857$\pm$0.007 & 10.989$\pm$ 0.02& 10.892$\pm$0.033 & 10.851$\pm$0.029  & 1.235 &           0.016  \\      
 14 &  341.736972  &58.117806  &0.562$\pm$0.015  &13.143$\pm$0.010 & 12.125$\pm$0.028& 12.021$\pm$0.035 & 11.897$\pm$0.028  & 5.463 &           0.014 \\       
 15 &  341.801389  &58.109528  &1.465$\pm$0.037  &16.999$\pm$0.044 & 14.196$\pm$0.034& 13.604$\pm$0.038 & 13.471$\pm$0.042  & 9.720 &           0.059 \\       
 16 &  341.737417  &58.104556  &0.659$\pm$0.014  &13.774$\pm$0.010 & 12.562$\pm$0.023& 12.429$\pm$0.032 & 12.347$\pm$0.028  & 0.406 &           0.011 \\       
 17 &  341.680750  &58.096278  &1.876$\pm$0.017  &15.996$\pm$0.022 & 12.603$\pm$0.023& 11.802$\pm$0.028 & 11.555$\pm$0.021  & 0.067 &           0.023 \\       
 18 &  341.938194  &58.094194  &1.270$\pm$0.024  &15.759$\pm$0.022 & 13.246$\pm$0.029& 12.772$\pm$0.036 & 12.564$\pm$0.03  & 2.552 &           0.036  \\      
 19 &  341.707417  &58.223417  &1.794$\pm$0.014  &15.534$\pm$0.020 & 12.294$\pm$0.024& 11.586$\pm$ 0.03 & 11.389$\pm$0.025  & 3.300 &           0.028 \\       
 20 &  341.707139  &58.221000  &1.775$\pm$0.012  &15.242$\pm$0.018 & 11.988$\pm$0.024& 11.337$\pm$ 0.03 & 11.077$\pm$0.023  & 0.767&           0.027 \\        
 21 &  341.710583  &58.140278  &4.099$\pm$0.010  &15.614$\pm$0.020 &  8.644$\pm$0.019&  7.563$\pm$0.033 &  7.135$\pm$0.021  & 1.050/20.71&      0.068 \\       
 22 &  341.759722  &58.124917  &3.678$\pm$0.012  &15.560$\pm$0.020 &  9.351$\pm$0.023&  8.180$\pm$0.038 &  7.743$\pm$0.016  & 48.879/1.02&     0.058 \\        
 23 &  341.865917  &58.123000  &-                &18.463$\pm$0.128 & 15.767$\pm$0.09 & 15.421$\pm$0.148 & 14.541            & 0.050&            0.189 \\       
 24 &  341.850444  &58.116472  &0.981$\pm$0.034  &16.470$\pm$0.030 & 14.385$\pm$0.039& 14.165$\pm$0.044 & 14.008$\pm$0.066  & 0.073&            0.044 \\       
 25 &  341.692778  &58.094972  &2.050$\pm$0.019  &16.490$\pm$0.032 & 12.756$\pm$0.023& 11.897$\pm$ 0.03 & 11.638$\pm$0.019  & 0.134&            0.046 \\       
 26 &  342.024056  &58.076861  &0.887$\pm$0.129  &18.334$\pm$0.147 & 16.334$\pm$0.127& 15.748$\pm$0.152 & 13.953           & 0.217&            0.205 \\       
 27 &  341.899417  &58.075500  &1.579$\pm$0.027  &16.713$\pm$0.037 & 13.794$\pm$0.051& 13.115$\pm$0.056 & 12.893$\pm$0.046  & 1.642&            0.074 \\       
 28 &  342.001861  &58.052222  &1.435$\pm$0.120  &18.912$\pm$0.156 & 16.374$\pm$0.118& 15.956$\pm$0.171 & 15.728$\pm$0.3 & 0.431&            0.202 \\        
 29 &  341.946583  &58.025083  &0.842$\pm$0.015  &13.568$\pm$0.015 & 12.139$\pm$0.025& 11.805$\pm$ 0.03 & 11.786$\pm$0.025  & 0.376&            0.022  \\      
 30 &  341.842972  &58.022194  &2.129$\pm$0.012  &14.895$\pm$0.016 & 10.797$\pm$0.031&  9.763$\pm$ 0.03 &  8.847$\pm$0.021  & 19.109&           0.049 \\       
 31 &  341.812083  &58.109389  &0.472$\pm$0.015  &12.673$\pm$0.008 & 11.851$\pm$0.025& 11.761$\pm$0.028 & 11.766$\pm$0.025  & 0.197&            0.012 \\       
 32 &  341.819500  &58.089972  &0.839$\pm$0.009  &14.064$\pm$0.020 & 12.506$\pm$0.026& 12.344$\pm$0.032 & 12.221$\pm$0.026  & 0.254&            0.032 \\       
 33 &  341.627417  &58.051194  &1.345$\pm$0.037  &16.825$\pm$0.045 & 14.414$\pm$0.041& 14.013$\pm$0.052 & 13.750$\pm$0.059  & 0.568&            0.106 \\       
 34 &  341.917028  &58.150917  &0.976$\pm$0.012  &13.678$\pm$0.010 & 11.846$\pm$0.023& 11.505$\pm$0.026 & 11.217$\pm$0.023  & 0.639&            0.018 \\       
 35 &  341.933778  &58.014139  &1.430$\pm$0.045  &17.615$\pm$0.070 & 15.171$\pm$0.049& 14.836$\pm$0.074 & 14.703$\pm$0.112  & 39.240&           0.087  \\      
 36 &  341.953778  &58.220056  &0.969$\pm$0.012  &13.907$\pm$0.010 & 12.247$\pm$0.03 & 11.815$\pm$0.036 & 11.729$\pm$0.029  & 0.197&            0.010  \\      
 37 &  341.956833  &58.219639  &1.196$\pm$0.027  &16.353$\pm$0.029 & 14.241$\pm$0.039& 13.908$\pm$ 0.05 & 13.761$\pm$0.056  & 0.334&            0.030  \\      
 38 &  341.779528  &58.184500  &3.140$\pm$0.012  &15.245$\pm$0.017 &  9.769$\pm$0.023&  8.674$\pm$0.031 &  8.256$\pm$0.023  & 24.676/1.039&           0.029 \\ 
 39 &  341.898806  &58.153333  &0.884$\pm$0.027  &15.892$\pm$0.022 & 14.157$\pm$0.035& 13.319$\pm$0.038 & 12.238$\pm$0.028  & 48.557&           0.038 \\       
 40 &  341.915639  &58.141833  &2.139$\pm$0.044  &17.966$\pm$0.078 & 14.092$\pm$0.035& 12.948$\pm$0.032 & 12.169$\pm$0.029  & 21.607&           0.133  \\      
 41 &  341.998556  &58.141056  &1.725$\pm$0.009  &12.434$\pm$0.010 &  9.293$\pm$0.023&  8.571$\pm$0.031 &  8.360$\pm$0.02  & 0.125&            0.008  \\      
 42 &  341.809667  &58.116389  &0.743$\pm$0.024  &15.506$\pm$0.017 & 13.761$\pm$0.025& 13.051$\pm$0.032 & 12.252$\pm$0.025  & 60.543&           0.026  \\      
 43 &  341.821583  &58.077361  &4.215$\pm$0.010  &15.829$\pm$0.020 &  8.687$\pm$0.026&  7.592$\pm$0.031 &  7.121$\pm$0.021  & 11.924&           0.031  \\      
 44 &  341.929639  &58.062694  &1.865$\pm$0.017  &16.073$\pm$0.025 & 12.642$\pm$0.025& 12.060$\pm$0.028 & 11.727$\pm$0.024  & 30.897/1.029&           0.070 \\  
 45 &  341.988722  &58.127778  &1.190$\pm$0.014  &15.193$\pm$0.015 & 13.186$\pm$0.027& 12.613$\pm$0.031 & 12.540$\pm$0.031  & 16.423&           0.016 \\       
 46 &  341.644250  &58.121667  &1.357$\pm$0.037  &17.043$\pm$0.047 & 14.349$\pm$0.032& 13.463$\pm$0.033 & 12.799$\pm$0.029  & 5.586/45.572&           0.059 \\  
 47 &  342.040500  &58.115694  &0.724$\pm$0.012  &13.388$\pm$0.012 & 12.042$\pm$0.021& 11.865$\pm$0.027 & 11.764$\pm$0.024  & 0.544&            0.017  \\      
 48 &  341.706528  &58.010278  &1.028$\pm$0.026  &15.627$\pm$0.026 & 13.837$\pm$0.028& 13.439$\pm$0.033 & 13.327$\pm$0.037  & 15.467&           0.029 \\       
 49 &  341.825194  &58.144667  &0.488$\pm$0.015  &11.989$\pm$0.010 & 11.191$\pm$0.023& 11.092$\pm$0.027 & 11.086$\pm$0.02  & 0.166&            0.015 \\       
 50 &  341.725167  &58.084167  &0.580$\pm$0.016  & 8.629$\pm$0.018 &  7.741$\pm$0.027&  7.718$\pm$0.055 &  7.682$\pm$0.016  & 1.057&         0.052  \\         
 51&   341.800278 &58.176111   &2.060$\pm$0.256  &19.215$\pm$0.318 & 15.866$\pm$0.078& 14.900$\pm$0.064 & 14.609$\pm$0.096  &46.365&         0.325 \\          
 52&   341.925278 &58.151139   &-                &-                & 15.306$\pm$0.081& 14.231$\pm$0.072 & 13.630$\pm$0.059  &16.845&         0.343 \\           
 53&   342.027278 &58.077194   &2.320$\pm$0.054  &17.957$\pm$ 0.128& 13.978$\pm$0.062& 12.636$\pm$0.062 & 11.651$\pm$0.039  &53.644&         0.372 \\          
 54&   341.770194 &58.100722   &0.529$\pm$0.007  &11.296$\pm$0.006 & 10.366$\pm$0.024& 10.236$\pm$ 0.03 & 10.050$\pm$0.021  &7.715& 0.019 \\                   
 55&   341.802167 &58.144944   &0.381$\pm$0.010  &10.656$\pm$0.014 & 10.005$\pm$0.023& 10.005$\pm$0.027 &  9.996$\pm$0.02  &0.143& 0.029 \\                                   
 56&   341.901778 &58.187639   &0.523$\pm$0.006  &10.354$\pm$0.007 &  9.421$\pm$0.022&  9.197$\pm$0.028 &  9.180$\pm$0.02  &0.394& 0.020 \\                                    
 57&   342.067833 &58.013167   &0.709$\pm$0.007  &10.404$\pm$0.007 &  9.140$\pm$0.022&  8.976$\pm$0.028 &  8.876$\pm$0.02  &0.200& 0.079 \\                                            
\hline                                                  
\end{tabular}                                           
\end{table*}                                            
%*************************************************************************************************************************************************
\begin{table}                                           
\caption{Proper motion data taken form Chen et al. (2011).}
\scriptsize                                             
\begin{tabular}{lllll}                                  
\hline
id&$\mu_{RA}$       &$\mu_{DEC}$      &$\mu_{RA}$        &$\mu_{DEC}$ \\
  & (SHAO)          & (SHAO)         & (UCAC3)         &(UCAC3) \\
   &(mas/yr)           &(mas/yr)     &     (mas/yr)    &         (mas/yr) \\
\hline
4  & -3.309$\pm$0.484 & -3.049$\pm$1.056 &  -8.1$\pm$1.0 &   -7.6$\pm$1.5 \\ 
6  &  3.253$\pm$0.730 & -1.002$\pm$1.314 &   0.2$\pm$2.3 &  -10.3$\pm$4.6 \\  
7  &  5.610$\pm$1.025 & -1.575$\pm$1.304 &  -5.7$\pm$2.7 &    1.2$\pm$5.9 \\  
8  &  2.731$\pm$0.532 & -0.508$\pm$1.787 &   2.0$\pm$3.4 &   -4.7$\pm$11.1 \\ 
11 &  1.084$\pm$0.405 &  0.259$\pm$1.445 &  -2.8$\pm$1.2 &   -3.1$\pm$2.2 \\ 
12 &  1.070$\pm$0.433 & -0.877$\pm$1.055 &  -0.5$\pm$3.3 &   -4.1$\pm$1.9 \\ 
13 & -0.287$\pm$0.382 & -1.353$\pm$1.415 &  -3.1$\pm$0.8 &   -3.8$\pm$0.9 \\ 
14 &  0.816$\pm$0.415 & -3.521$\pm$0.682 &  -2.8$\pm$1.0 &   -3.3$\pm$1.8 \\ 
29 &-14.553$\pm$0.532 &-14.721$\pm$1.050 & -20.4$\pm$2.7 &  -15.7$\pm$3.1 \\ 
31 &  0.848$\pm$0.373 & -2.625$\pm$0.789 &  -3.6$\pm$2.1 &   -4.1$\pm$1.1 \\ 
32 &  2.735$\pm$0.583 & -7.836$\pm$1.718 &  -6.5$\pm$2.0 &   -5.0$\pm$6.1 \\ 
34 & -1.160$\pm$0.641 & -6.407$\pm$1.499 &  -3.5$\pm$2.0 &   -4.6$\pm$6.2 \\ 
36 & 12.446$\pm$0.607 & -1.615$\pm$1.890 &   9.2$\pm$2.6 &   -4.9$\pm$4.1 \\ 
41 &  2.392$\pm$0.726 & -3.801$\pm$1.583 &  -1.6$\pm$2.0 &   -0.7$\pm$1.8 \\ 
47 & -2.893$\pm$0.655 & -3.094$\pm$1.961 &  -3.9$\pm$1.4 &   -2.6$\pm$4.6 \\ 
49 &  1.278$\pm$0.426 & -1.676$\pm$1.181 &  -2.4$\pm$0.6 &   -3.6$\pm$0.7 \\ 
50 &  -4.11$\pm$1.290  & -2.80$\pm$1.190  &  -2.7        &       -3.6      \\
54 & -0.342$\pm$0.692 & -1.161$\pm$0.813 &  -2.3$\pm$0.8 &   -3.5$\pm$0.8 \\ 
55 &  0.292$\pm$0.417 & -1.658$\pm$1.848 &  -1.8$\pm$0.6 &   -2.9$\pm$0.6 \\ 
\hline
\end{tabular}
\end{table}

\begin{table}
\caption{X-ray data.}
\scriptsize
\begin{tabular}{llll}
\hline
id&PNf       &Mf      &Avef    \\
\hline
7 &   9.36&  10.08&   9.72 \\
13&   3.98&   4.79&    4.39 \\
15&   1.58&   1.36&   1.47 \\  
27&   2.54&   1.57&   2.06 \\  
30&    3.5&   3.72&   3.61 \\  
50&   83.9&  80.65&  82.28 \\  
53&    22.94&  28.46&  25.7 \\ 
\hline
\end{tabular}

PNf: flux from PN  detector in units of 10$^{-14}$ erg/s/cm$^2$ \\
Mf: flux from MOS detector in units of 10$^{-14}$ erg/s/cm$^2$ \\
Avef: Average flux from PN and MOS detector in units of 10$^{-14}$ erg/s/cm$^2$ \\
\end{table}

%*************************************************************************************************************************************************
%*************************************************************************************************************************************************
\begin{figure}
\includegraphics[width=9cm]{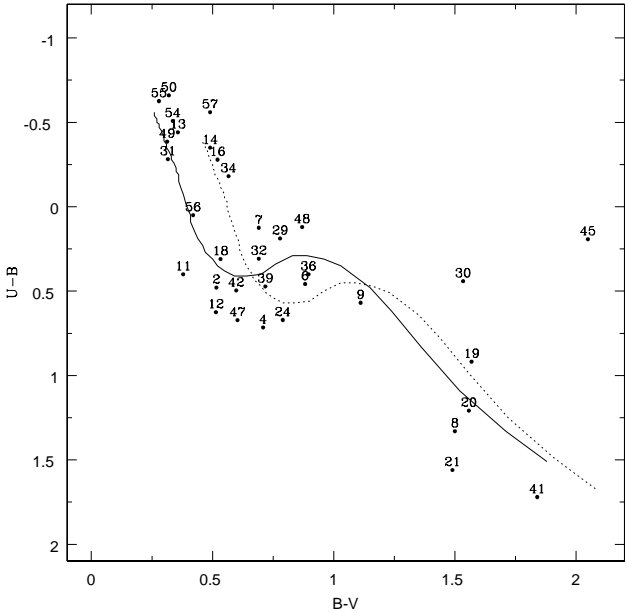}
\caption{$U-B/B-V$ two colour diagram for identified variable stars. The $UBV$ data have been taken from 
Chen et al. (2011). The solid curve represents the zero-age-main-sequence (ZAMS) by Girardi et al. (2002) shifted along the reddening vector of 0.72 for $E(B-V)= 0.5$ mag and $E(B-V)= 0.7$ mag. }
\end{figure}
%*************************************************************************************************************************************************

%*************************************************************************************************************************************************
\begin{figure}
\includegraphics[width=9cm]{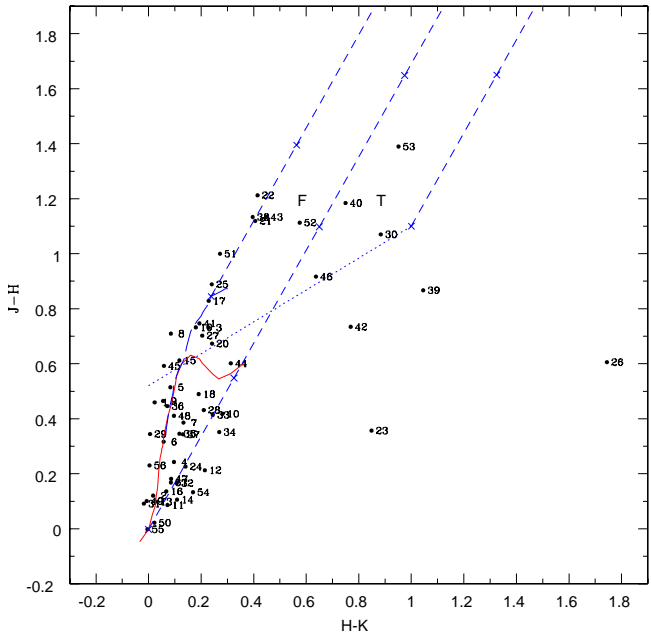}
\caption{($J-H/H-K$) TCD for variable stars lying in the field of
 NGC 7380. $JHK$ data have been taken from 2MASS catalogue (Cutri et al. 2003). The sequences for dwarfs (solid curve) and giants (long dashed curve) are from Bessell \& Brett (1988). The dotted curve represents the locus of TTSs (Meyer et al. 1997).
The small dashed lines represent the reddening vectors (Cohen et al. 1981). 
 The crosses on the reddening vectors represent an increment of visual extinction of $A_{V}$ = 5 mag. } 
\end{figure}
%*************************************************************************************************************************************************

%*************************************************************************************************************************************************
\begin{figure}
\includegraphics[width=9cm]{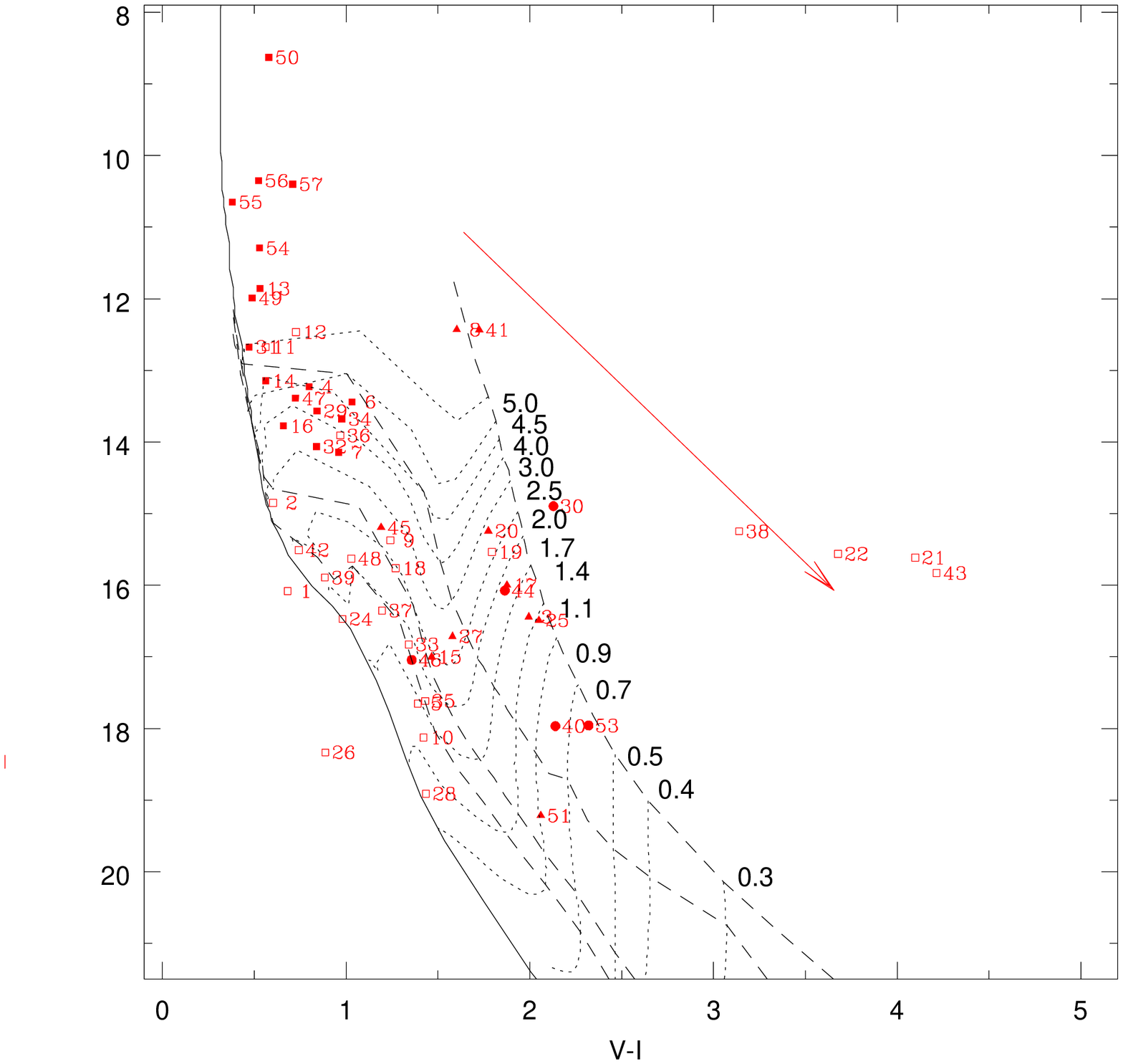}
\caption{$V/V-I$ colour-magnitude diagram of the cluster NGC 7380 for variable candidates. The filled circles and triangles represent probable CTTSs and WTTSs, respectively, whereas filled squares and open squares represent MS population in the cluster region and field population towards the cluster region, respectively. 
The ZAMS by Girardi et al. (2002)
 and PMS isochrones for
0.1, 1, 5, 10 Myrs by Siess et al. (2000) are shown.
The dotted curves show PMS evolutionary tracks of stars of different masses.
The isochrones and evolutionary tracks  are corrected for the cluster distance and $E(V-I)=0.625$ mag. The straight line indicates reddening vector for A${_V}/E(V-I)=2.48$. 
}
\end{figure}
%*************************************************************************************************************************************************

\section{Association of Young stellar Objects (YSOs) with the cluster}
To understand the nature of the variable candidates it is necessary to find out their association to the
cluster NGC 7380. The $U-B/B-V$ and $J-H/H-K$ TCDs, $V/V-I$ CMD have been used to find out the association of the identified 
variables with NGC 7380 region. In addition, we have also used proper motion data from Chen et al. (2011) listed  in Table 2 as well as X-ray data listed in Table 3 to identify the PMS variables associated with NGC 7380
region.

Since PMS stars (both classical TTSs and weak-line TTSs) are strong X-ray emitters, the X-ray observations of PMS stars play an important
role in identification of PMS stars associated with star forming regions
(see e.g., Pandey et al 2014). In fact, the level of X-ray emission in the PMS stars, which is higher than that of MS stars, provides a very efficient
mean of selecting PMS stars.
However, the origin of X-ray emission from PMS low mass stars is still poorly understood.
In massive stars, the X-ray emission arises from shocks in radiatively-driven winds (Lucy \& White 1980; Owocki \& Cohen 1999; Kudritzki \& Puls 2000; Crowther 2007), while in the low-mass stars, rotation with convective envelopes drives a magnetic dynamo leading to strong X-ray emission (Vaiana et al. 1981; G\"udel 2004). Intermediate mass stars, on the other hand, are expected to be X-ray dark because (a) the wind is not strong enough to produce X-rays as in the case of massive stars (see Lucy \& White 1980; Kudritzki \& Puls 2000), and (b) being fully radiative internal structure, the dynamo action cannot support the X-ray emission.

The X-ray data for the NGC 7380 region have been taken from Bhatt et al. (2010, 2013). They have  analyzed archival X-ray data from XMM-Newton observations for the region containing
NGC 7380. Data were acquired simultaneously with EPIC-PN camera (Str\"uder et al. 2001) and two nearly identical EPIC-MOS (MOS1 and MOS2; Turner et al. 2001) cameras.  
We identified the X-ray data for seven variable stars namely 7, 13, 15, 27, 30, 50 (DH cep) and 53. The matching radius
was 6 arcsec. 

To establish membership further we used proper motion data of stars in NGC 7380 from Chen et al. (2011). Out of 57  stars only 18 stars have proper motion data. 
Table 2 lists two data sets of proper motions and associated errors from UCAC3 and SHAO.
Typical errors in proper motions range from 1 to 10 mas/yr for UCAC3, and $<$ 1 mas/yr for the SHAO. The mean proper motion for NGC 7380 has been determined
 as $\mu_{RA}$=$-1.74\pm0.84$ and $\mu_{DEC}$=$-2.52\pm0.81$ by Baumgardt et al. (2000).
The proper motion values of the present sample except star nos 7, 29, 32 and 36  are within 3$\sigma$ limit.
Thus proper motion data suggest that star nos 7, 29, 32 and 36 could be non-members.

In the absence of kinematical data of the remaining stars  we have used photometric data to determine the their membership to the cluster. We have used $(U-B)/(B-V)$ and $(J-H)/(H-K)$ TCDs and $V/V-I$ CMD to find out the probable member stars associated with the NGC 7380 region. 

\subsection{$(U-B)/(B-V)$ and $(J-H)/(H-K)$ TCDs}
Fig. 5 shows $(U-B)/(B-V)$ TCD for the variable candidates. The $UBV$ data have  been taken from 
Chen et al. (2011) and WEBDA. Out of 53 variable candidates UBV data are available for 33 stars and these are plotted in  $(U-B)/(B-V)$ TCD. The distribution of 
variables in Fig. 5 reveals a variable reddening in the cluster region. 
 The identification of young stellar objects (YSOs) or estimation of reddening
for YSOs is not possible using the $(U-B)/(B-V)$ TCD because the $U$ and $B$ 
band fluxes may be affected by excess due to accretion.
The sources lying near the MS with O to A spectral type are considered as
MS type stars.
  
The young stellar objects (YSOs) are easily recognized by
their H${\alpha}$
emission, NIR excess, or X-ray emission, therefore $(J-H)/(H-K)$ TCD is one of the very useful tools to identify PMS objects. Fig. 6 shows the $(J-H)/(H-K)$ TCD for variable stars.  
We have used $JHK$ data from the 2MASS catalogue (Cutri et al. 2013) and converted to CIT system using the relations given in the 2MASS website (http://www.astro.caltech.edu/~jmc/2mass/v3/transformations/).  The solid
and long dashed lines in Fig. 6 represents unreddened MS and giant locus (Bessell \& Brett 1988) respectively.
The dotted line indicates the intrinsic locus of CTTSs (Meyer et al. 1997).
The parallel dashed lines are the reddening vectors drawn from the tip (spectral type M4) of the giant branch (`left reddening line'), from the base (spectral type A0) of the MS branch (`middle reddening line') and from the tip of the intrinsic Classical T Tauri Stars line (`right reddening line'). The extinction ratios $A_J/A_V= 0.265, A_H/A_V= 0.155$ and $A_K/A_V= 0.090$ have been adopted from Cohen et al. (1981).
The sources lying in `F' region could be
either field stars (MS stars, giants), Class III or Class II sources
with small NIR excesses. The sources lying in the `T' region e.g. 30, 40, 46 and 53 are considered to be
mostly CTTSs (Class II objects).
Majority of the variable candidates are found to be distributed below the intrinsic locus of CTTSs and
these should be B type MS stars. 
The stars numbered 3, 10, 20, 21, 27, 41, 43 and 52 located in the `F' region
and lie above the extension of CTTSs locus could be the WTTSs (Class III sources).
 
\begin{figure}
\includegraphics[width=7cm]{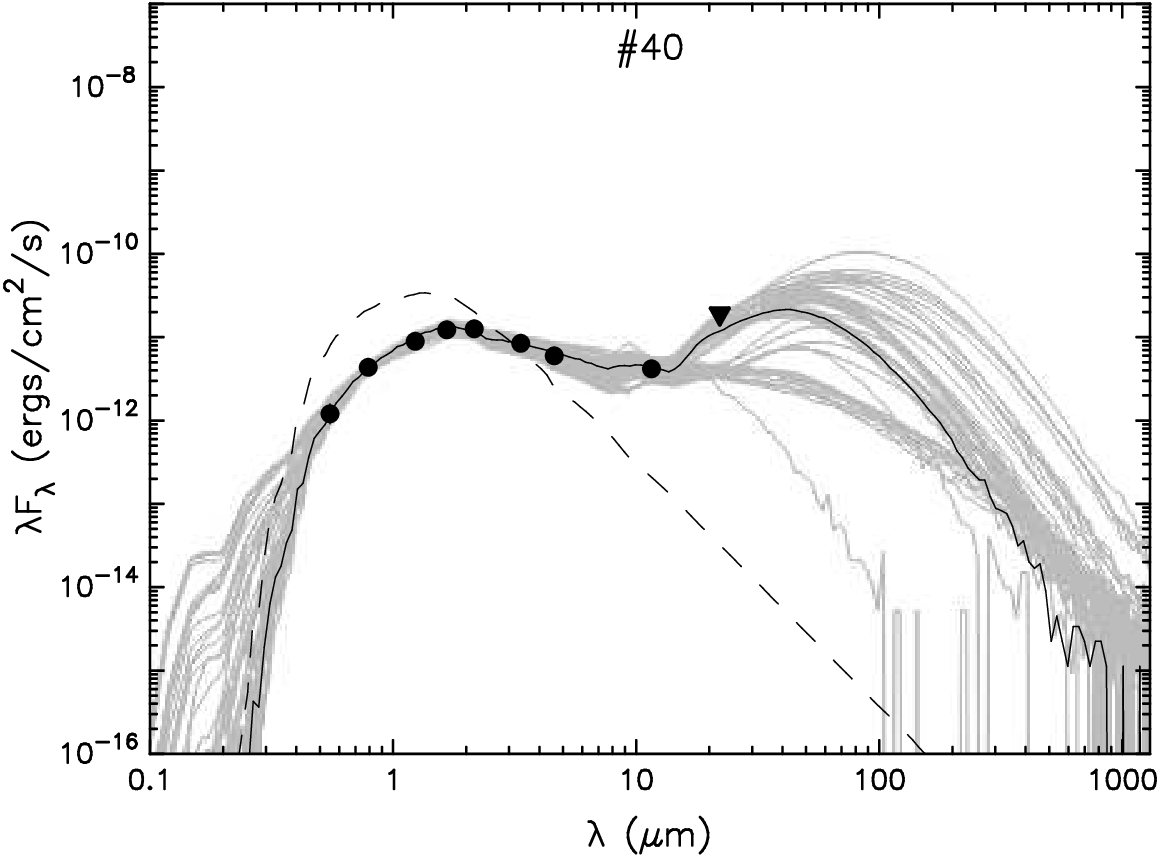}
\includegraphics[width=7cm]{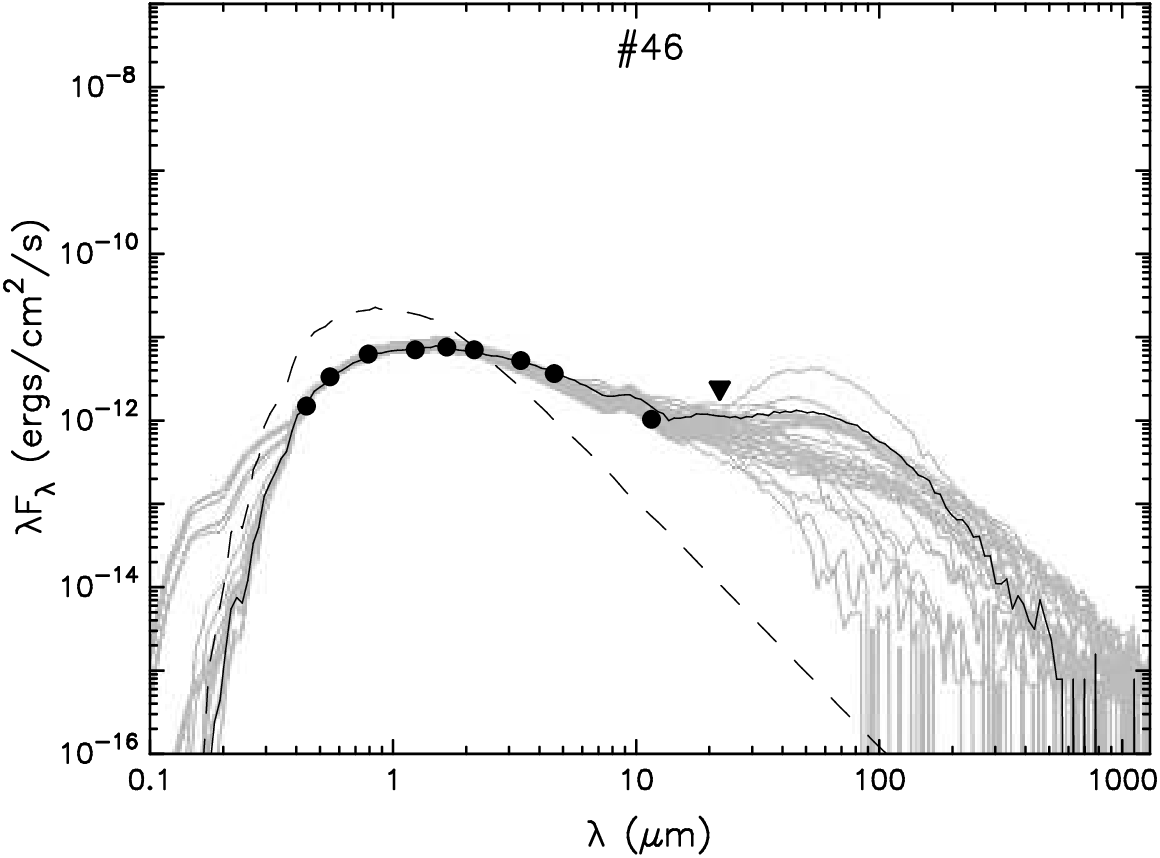}
\caption{SEDs of the YSOs ID 40 and ID 46.
The black line shows the best fit model, and the grey lines show subsequent models
that satisfy $\chi^2 - \chi^2_{\rm min} \leq 2N_{\rm data}$ criteria. The
dashed line shows the stellar photosphere corresponding to the
central source of the best fitting model. The circles denote the
 observed flux values. The triangle represents 22 $\micron$ flux which is considered as upper limit while fitting.}
\end{figure}

\subsection {$V/V-I$ colour-magnitude diagram}
$V/V-I$ CMD is also an additional tool to identify probable  member of the cluster. Fig. 7 shows the $V/V-I$ CMD for all the identified variables in the region. 
The filled circles and triangles represent probable CTTSs and WTTSs, respectively, whereas filled squares and open squares represent MS population in the cluster region and field population towards the cluster region, respectively identified on the basis of
$U-B/B-V$ and $J-H/H-K$ TCDs. 
In Fig. 7 we have also plotted theoretical isochrone for 4 Myr for Z=0.02 (continuous line) by Girardi et al. (2002) and PMS isochrones for various ages and evolutionary tracks for various masses by
Siess et al. (2000). All the isochrones and evolutionary tracks are corrected for the distance (2.6 kpc) and minimum reddening $E(V-I)$=0.625 mag. The minimum value of $E(V-I)$ has been estimated using the relation $E(V-I)/E(B-V)=1.25$ and $E(B-V)=0.50$ mag. 

 The association of variables to the cluster NGC 7380 is marked `yes' on the basis of above criteria as well as
classification of variables is marked in Table 4. The variables not associated to the cluster is marked `no' in Table 4.

\subsection{Field star contamination}
To estimate the field star contribution towards the direction of the cluster NGC 7380 we generated star counts in 1$\times$1 degree${^2}$ field using the  Besancon Galactic model of stellar population synthesis (Robin et al. 2003).
The distance to the cluster and $E(B-V)$ has been taken as 2.6 kpc and 0.5-0.7 mag respectively.
The model simulations with distance $<$ 2.6 kpc and $A_{V}$ $\leq$ 1.55 mag give the foreground contamination, and that with distance $>$ 2.6 kpc and $A_{V}$ $\geq$ 2.17 mag give the background population. The advantage of this method is that we can separate the foreground (distance $<$ 2.6 kpc) and the background (distance $>$ 2.6 kpc) field star contamination.
The fraction of contaminating stars upto $V\sim$18 has been determined as number of foreground and background stars (estimated from the model) over total number stars present in the observed
cluster region (estimated from observed data) and it comes out to be 36\%.
Out of 57 variables studied here, 24 are found to be confirmed field star population which gives a contamination of about 42\%, which is comparable with the model prediction.

\section{Age and mass estimation}
The age and mass of a PMS variable can be estimated with the help of PMS evolutionary tracks. The estimated ages of the majority of the YSOs
are in the range of 0.1 to 5 Myr which are comparable to the ages of TTSs. The masses of the majority of PMS variables range from $\sim$ 0.6 to $\sim$ 2.30 $M_{\odot}$. 
The associated errors in determination of age and mass can be of two kinds; random errors in observations and systematic errors due to the use of different theoretical evolutionary tracks. We have estimated the effect of random errors in determination of age and mass by propagating the random errors to the observed estimations of $V, V-I$ and $E(V-I)$ by assuming normal error distribution and using the Monte Carlo simulations (see e.g., Chauhan et al. 2009).
Since we have used model by Siess et al. (2000) only for all the PMS stars, present age and mass estimations are not affected by the systematic errors.
The presence of binaries may be another source of errors. The presence of binary will brighten a star, consequently yielding a younger age. In the case of equal mass binary, we expect an error of $\sim$50 to 60\% in age estimation of the PMS stars. However, it is difficult to estimate the influence of binaries on mean age estimation as the fraction of binaries is not known.
The estimated ages and masses along with their errors are given in Table 5.

\section{Spectral Energy Distribution}
To characterize the probable YSO variables further we have analyzed the 
spectral energy distribution of these sources using the radiative transfer models by Robitaille et al. (2006, 2007).
 Interpreting SEDs using the radiative transfer code
is subject to degeneracy, however spatially resolved multiwavelength
observations can reduce the degeneracy. We compiled the data of the YSOs at optical ($BVI$; present work), NIR ($JHK$; 2MASS survey), 
and WISE (3.4, 4.6, 12.0, and 22.0 $\micron$; Cutri et al. 2012) bands, wherever available. We used 22 $\micron$
data as upper-limit owing to its large beam ($\sim$22 arcsec) and crowded
nature of the field. While fitting models we scaled the SED models to the distance of the cluster (i.e., 2.6 $\pm$ 0.4 kpc) and allowed
a maximum $A_{V}$ value determined by tracing back the YSOs’ current location on $J/J-H$ diagram to the intrinsic dwarf locus
along the reddening vector (see e.g., Samal et al. 2010), where
the minimum extinction adopted as 1.5 mag, which is the foreground extinction towards the direction of cluster.

Fig. 8 shows the SEDs of two sources as examples. The WISE data were not available for the stars 15 and 45. Due to lack of
mid and far-infrared, and millimeter data points, it is quite apparent that the SED models show high degree of degeneracy;
nonetheless, barring star 8 and 41 SEDs of majority sources indicate the presence of IR-excess emission, possibly emission
from circumstellar disk. We found that
the observed SEDs of two sources (IDs. 8 and 41) are not well
constrained. These two sources are possibly disk-less stars or
other reddened stellar sources along the line of sight.
It is not possible to characterize all the SED parameters from
the models due to limited observational data points. However, as
discussed by Robitaille et al. (2007), some of the parameters can
still be constrained depending on the available fluxes. In the present study, the SED
models between 1$\micron$ to 12$\micron$ represent the data fairly well
, hence, the stellar parameters are expected to be better
constrained. However, it is worth noting that precise determination of stellar parameters using SED models 
requires data from optical to millimeter bands. 
In Table 5, the
mass and age estimated using the SED model are also tabulated. Since our SED models are highly degenerate, the best-fit model is unlikely
to give an unique solution, the tabulated values are
the weighted mean and standard deviation of the physical parameters obtained from the models that satisfy
 $\chi^{2}$-$\chi^{2}_{min}$$\le$ 2N$_{data}$ data weighted by 
e$^{({{-\chi}^2}/2)}$ of each model as done in Samal et al. (2012), where  $\chi^{2}_{min}$ is the goodness-of-fit parameter for the best-fit model and $N_{\rm data}$
is the number of input observational data points.
Table 5 indicates that age and mass estimation using the SED models are higher by about 1.5 times 
in comparison to age and mass estimates using the $V/V-I$ CMD.
The estimate could be considered reasonable because of various uncertainties involved in
SED modeling and lack of longer-wavelength data points to fully
sample our SED.

\begin{table}
\caption{Classification of variable stars.}
\tiny
\begin{tabular}{ccccccc}
\hline
ID   &         $J-H/H-K$  &    $U-B/B-V$       &  $V/V-I$  &   Proper   &       X-ray    &   classification  \\
     &              &           &            &   motion   &                &                   \\
\hline
 1   &       no     &   -       &     no     &   -        &       -        &    Field          \\
 2   &       no     &   no      &     no     &   -        &       -        &    Field          \\
 3   &       yes    &   -       &     yes    &   -        &       -        &    PMS/WTTS          \\
 4   &       yes    &   yes     &     yes    &   yes      &       -        &    MS          \\
 5   &       no     &   -       &     no     &   -        &       -        &    Field          \\
 6   &       yes    &   yes     &     yes    &   yes      &       -        &    MS          \\
 7   &       yes    &   yes     &     yes    &   no       &       yes      &    MS/Field          \\
 8   &       ?    &   no      &     yes    &   yes      &       -        &    Field          \\
 9   &       no     &   no      &     no     &   -        &       -        &    Field          \\
 10  &       No     &   -       &     no     &   -        &       -        &    Field          \\
 11  &       no     &   no      &     no     &   yes      &       -        &    Field          \\
 12  &       no     &   no      &     no     &   yes      &       -        &    Field          \\
 13  &       yes    &   yes     &     yes    &   yes      &       yes      &    MS          \\
 14  &       yes    &   yes     &     yes    &   yes      &       -        &    MS          \\
 15  &       yes    &   -       &     ye     &   -        &       yes      &    PMS/WTTS          \\
 16  &       yes    &   yes     &     yes    &   -        &       -        &    MS          \\
 17  &       yes    &   -       &     yes    &   -        &       -        &    PMS/WTTS          \\
 18  &       no     &   no      &     no     &   -        &       -        &    Field          \\
 19  &       no     &   no      &     no     &   -        &       -        &    Field          \\
 20  &       yes    &   no      &     yes    &   -        &       -        &    PMS/WTTS          \\
 21  &       no     &   -       &     no     &   -        &       -        &    Field          \\
 22  &       no     &   -       &     no     &   -        &       -        &    Field          \\
 23  &       no     &   -       &     -      &   -        &       -        &    Field          \\
 24  &       no     &   no      &     no     &   -        &       -        &    Field          \\
 25  &       no     &   -       &     yes    &   -        &       -        &    PMS/WTTS          \\
 26  &       no     &   -       &     no     &   -        &       -        &    Field          \\
 27  &       yes    &   -       &     yes    &   -        &       yes      &    PMS/WTTS          \\
 28  &       no     &   -       &     no     &   -        &       -        &    Field          \\
 29  &       yes    &   yes     &     yes    &   no       &       -        &    MS/Field          \\
 30  &       yes    &   yes     &     yes    &   yes      &       yes      &    PMS/CTTS          \\
 31  &       yes    &   yes     &     yes    &   yes      &       -        &    MS          \\
 32  &       yes    &   yes     &     yes    &   no       &       -        &    MS/Field          \\
 33  &       no     &   -       &     no     &   -        &       -        &    Field          \\
 34  &       yes    &   yes     &     yes    &   no       &       -        &    MS          \\
 35  &       no     &   -       &     no     &   -        &       -        &    Field          \\
 36  &       no     &   no      &     no     &   yes      &       -        &    Field          \\
 37  &       no     &   -       &     no     &   -        &       -        &    Field          \\
 38  &       no     &   -       &     no     &   -        &       -        &    Field          \\
 39  &       no     &   no      &     no     &   -        &       -        &    Field          \\
 40  &       yes    &   -       &     yes    &   -        &       -        &    PMS/CTTS          \\
 41  &       ?      &   no      &     yes    &   yes      &       -        &    Field          \\
 42  &       no     &   no      &     no     &   -        &       -        &    Field          \\
 43  &       no     &   -       &     no     &   -        &       -        &    Field          \\
 44  &       yes    &   -       &     yes    &   -        &       -        &    PMS/WTTS          \\
 45  &       ?      &   no     &     yes     &   -        &       -        &    PMS/WTT/Field          \\
 46  &       yes    &   -       &     yes    &   -        &       -        &    PMS/CTTS          \\
 47  &       yes    &   yes     &     yes    &   yes      &       -        &    MS          \\
 48  &       no     &   no      &     no     &   -        &       -        &    Field          \\
 49  &       yes    &   yes     &     yes    &   yes      &       -        &    MS          \\
 50  &       yes    &   yes     &     yes    &   yes      &       yes      &    MS          \\
 51  &       no     &   -       &     yes    &   -        &       -        &    PMS/WTTS          \\
 52  &       yes    &   -       &     -      &   -        &       -        &    PMS/WTTS          \\
 53  &       yes    &   -       &     yes    &   -        &       yes      &    PMS/CTTS          \\
 54  &       yes    &   yes     &     yes    &   yes      &       -        &    MS          \\
 55  &       yes    &   yes     &     yes    &   yes      &       -        &    MS          \\
 56  &       yes    &   yes     &     yes    &   -        &       -        &    MS          \\
 57  &       yes    &   yes     &     yes    &   -        &       -        &    MS          \\
\hline
\end{tabular}
\end{table}
%*************************************************************************************************************************************************

%********************************************************************************************************************
\begin{table}
\scriptsize
\caption{Mass and age of probable PMS stars.}

\begin{tabular}{ccccc}
\hline
            
id &    mass($V/V-I$)&    age($V/V-I$) & mass(SED)&    age(SED) \\
   &    $M_{\odot}$ &   Myr& $M_{\odot}$ &Myr    \\
\hline
3  &     1.17$\pm$0.13& 0.25$\pm$0.02    &     3.262$\pm$0.467& 0.533$\pm$ 0.120      \\
15 &     1.92$\pm$0.16& 4.55$\pm$ 2.06   &       -            & -                     \\
17 &     1.58$\pm$0.11& 0.26$\pm$ 0.03   &     3.112$\pm$0.472&  0.627$\pm$0.201      \\
20 &     2.36$\pm$0.15& 0.23$\pm$ 0.03   &     3.011$\pm$0.006&  1.193$\pm$0.006       \\
25 &     1.07$\pm$0.06& 0.22$\pm$ 0.05   &     2.356$\pm$0.245&  0.401$\pm$0.135       \\
27 &     2.16$\pm$0.06& 1.60$\pm$ 0.46   &     2.188$\pm$0.254&  2.086$\pm$1.067       \\
30 &     2.06$\pm$0.08& 0.10$\pm$ 0.01   &     6.164$\pm$0.021&  0.262$\pm$0.004        \\
40 &     0.78$\pm$0.08& 0.59$\pm$ 0.06   &     1.622$\pm$1.055&  0.523$\pm$0.434        \\
44 &     1.58$\pm$0.11& 0.28$\pm$ 0.03   &     2.377$\pm$0.754&  0.485$\pm$0.164        \\
45 &     2.28$\pm$0.09& 4.22$\pm$ 0.57   &     -            & -                      \\
46 &        -         &$>$5.00           &     1.816$\pm$0.249&  4.885$\pm$1.005         \\
51 &     0.73$\pm$0.36& 4.00$\pm$ 4.47   &     1.135$\pm$0.373&  4.965$\pm$2.976         \\
52 &      -           &  -               &     -            & -                                           \\
53 &     0.62$\pm$0.05& 0.32$\pm$0.19    &     3.421$\pm$0.850&  0.107$\pm$0.053         \\
\hline
\end{tabular}
\end{table}

\begin{figure}
\includegraphics[width=9cm, height=8cm]{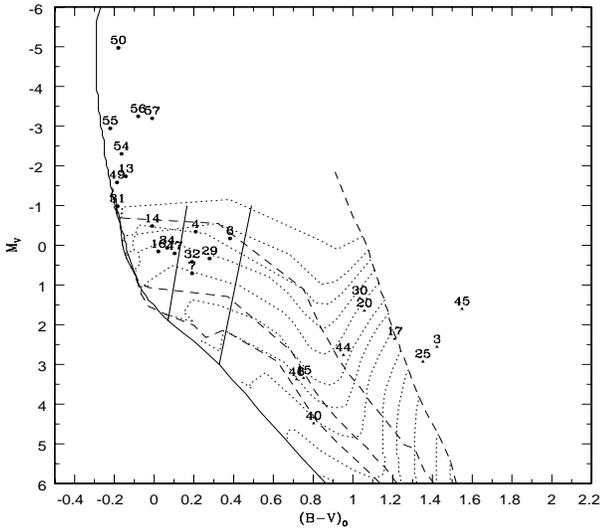}
\caption{$M_{V}/(B-V)_{0}$ colour-magnitude diagram for the MS and PMS stars of the cluster NGC 7380. The ZAMS by Girardi et al. (2002)
 and PMS isochrones for
0.1, 1, 5, 10 Myrs by Siess et al. (2000) are shown.
The dotted curves show PMS evolutionary tracks of stars of different masses. The theoretical instability strip of Cepheids is taken from the literature
(see Zwintz \& Weiss 2006).}
\end{figure}

\begin{figure}
\includegraphics[width=9cm]{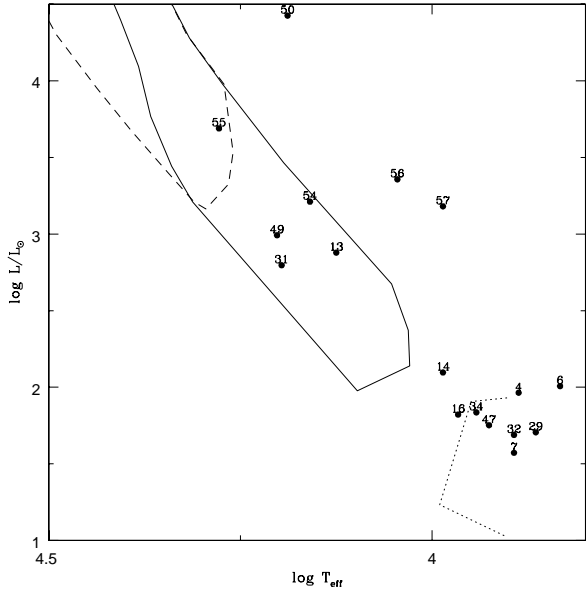}
\caption{ $\log(L/L_{\odot})/ \log T_{eff}$ diagram for the probable MS variables towards the cluster NGC 7380. The theoretical SPB instability strip (continuous curve),
 empirical $\delta$ Scuti instability strip (dotted curve) and the location of $\beta$ Cep stars (dashed curve) are taken from the literature (see Balona et al. 2011; references therein).
}
\end{figure}

%*************************************************************************************************************************************************
\begin{figure}
\includegraphics[width=7cm, height=9cm]{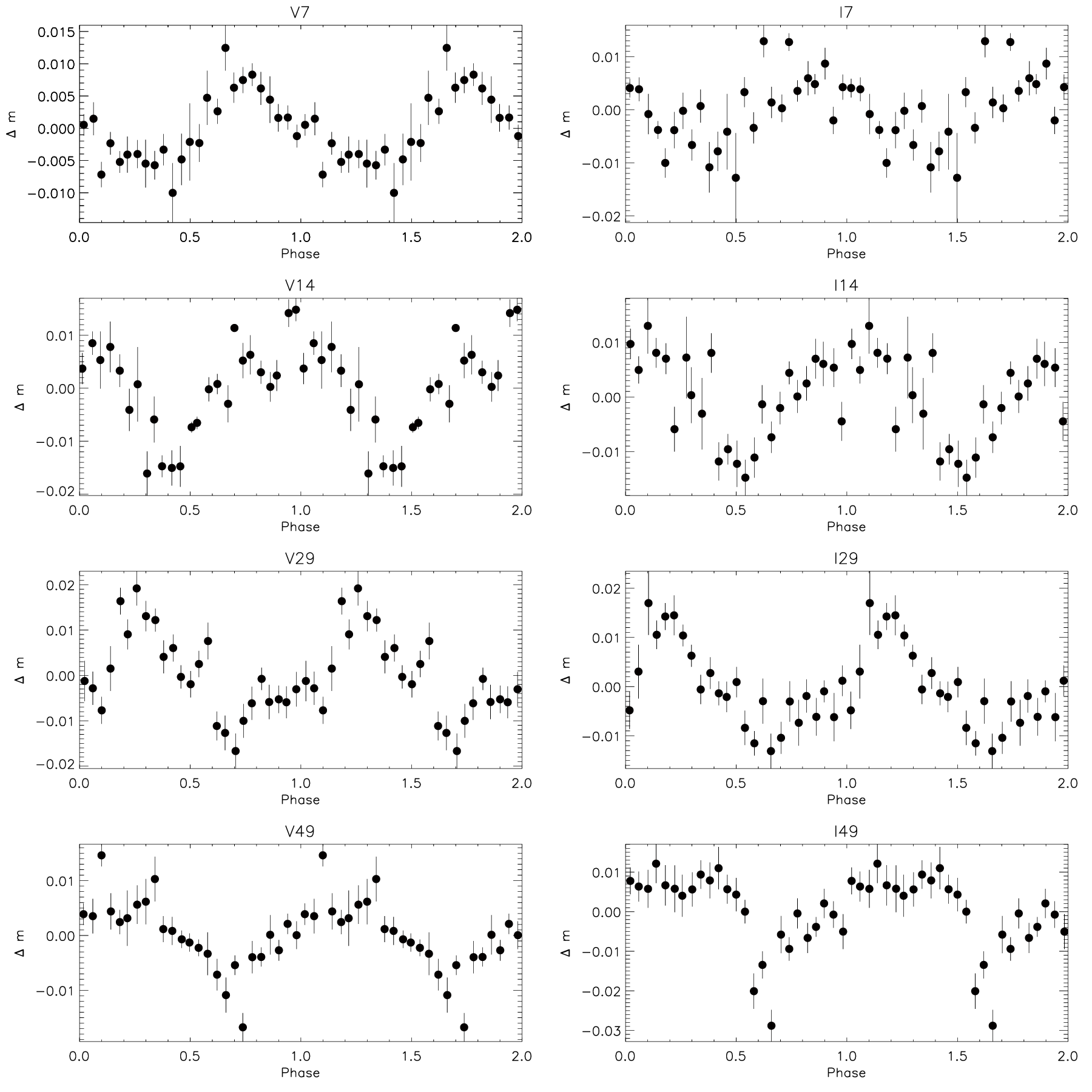}
\caption{The sample phased light curves of MS variable stars in $V$ and $I$ bands.  The complete figure is available electronic
form only.} 
\end{figure}

\begin{figure}
\includegraphics[width=7cm, height=9cm]{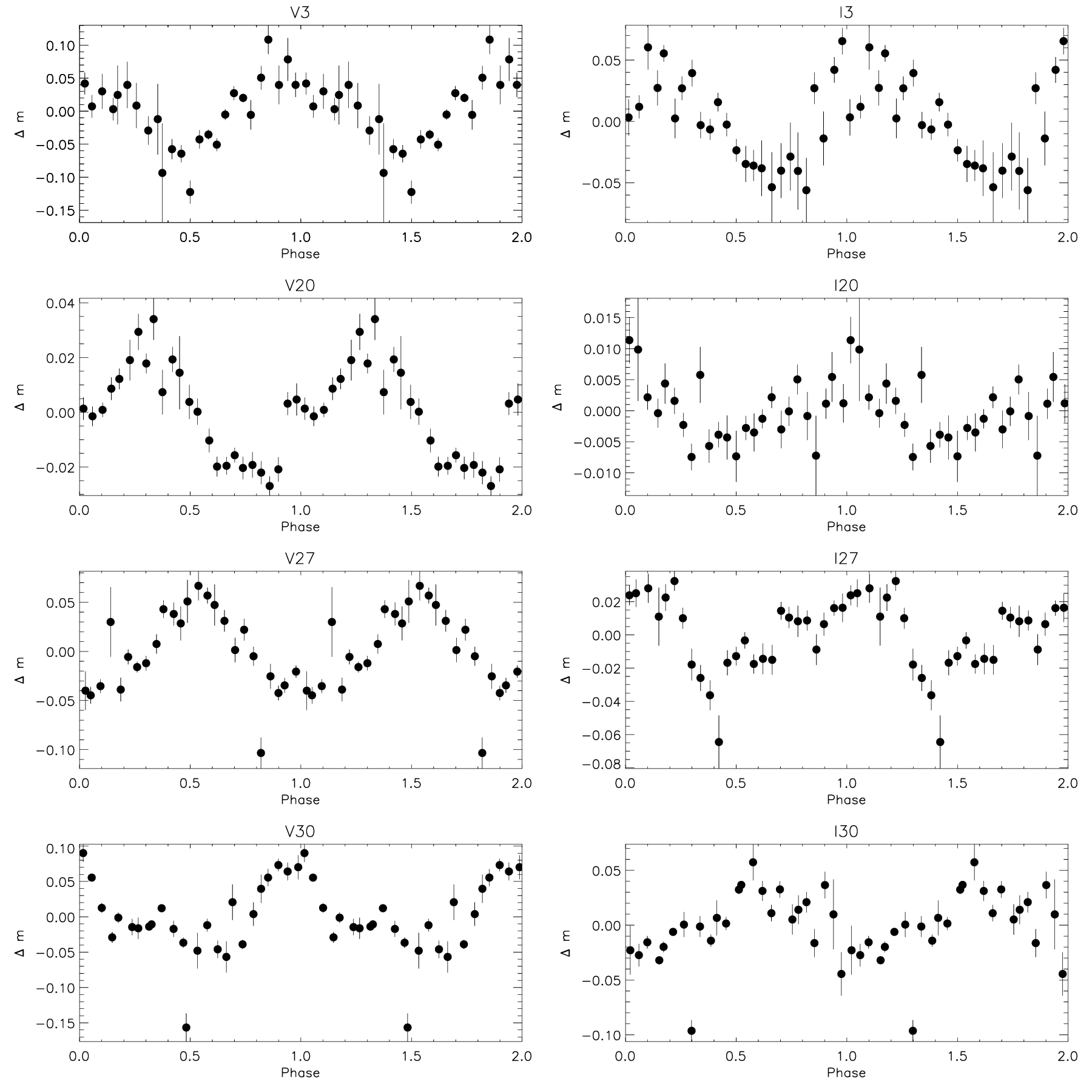}
\caption{The sample phased light curves of PMS variable stars in $V$ band and $I$ bands. The complete figure is available electronic
form only.} 
\end{figure}

\begin{figure}
\includegraphics[width=7cm, height=9cm]{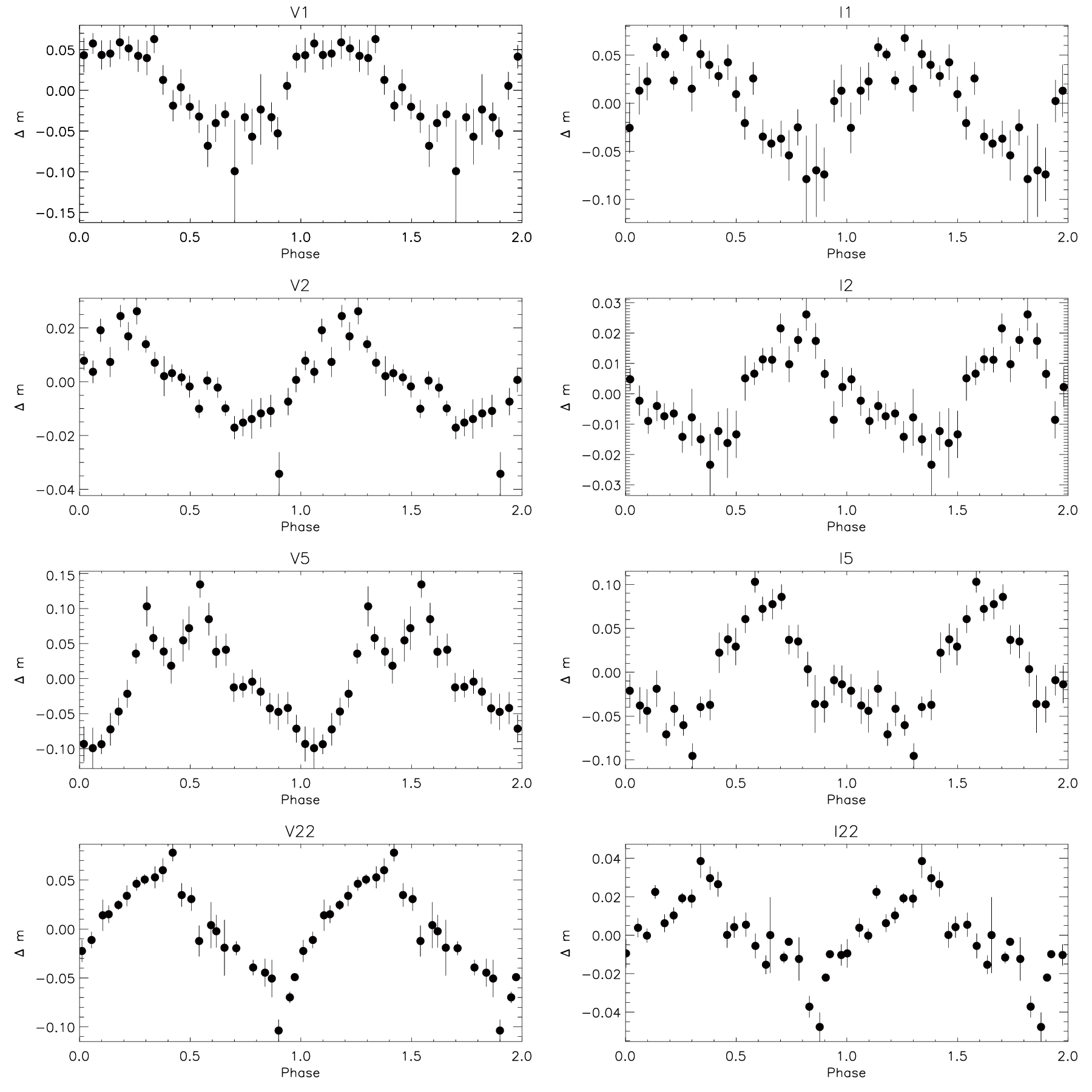}
\caption{The sample phased light curves of field variable stars in$V$ and $I$ bands.  The complete figure is available electronic
form only.} 
\end{figure}

%*************************************************************************************************************************************************
\section{$M_{V}/(B-V)_{0}$ and Luminosity $(L/L_{\odot})$ vs. Effective Temperature ($T _{eff}$) diagram}
The $M_{V}/(B-V)_{0}$ CMD for the identified MS and PMS stars is shown in Fig. 9. The intrinsic $(B-V)$ colours of MS stars have been
determined using the Q-method as described by Guti$\acute{e}$rrez-Moreno (1975).
For PMS stars we have used the average reddening ($E(B-V)$=0.5 mag) of the region. Fig. 10 shows the $\log(L/L_{\odot})/\log T_{eff}$ diagram for the MS variables.
The absolute magnitude $M_{V}$ was converted to luminosity using the relations $\log(L/L_{\odot})=-0.4(M_{bol}-M_{bol,\odot})$, and $M_{bol}=M_{V}+BC$, where BC is
the bolometric correction. The bolometric magnitude $M_{bol,\odot}$ for the Sun has been taken as 4.73 mag (Torres 2010).
To determine BC and effective temperature $T_{eff}$
we used the relations between $T_{eff}$-
intrinsic $(B-V)$ colours, and between  $T_{eff}$-BC by Torres (2010).
The luminosity ($\log L/L_{\odot}$), $M_{bol}$ and $\log T_{eff}$ and BC of MS stars
are listed in Table 6.

\begin{table}
\caption{The effective temperature ($T_{eff}$), bolometric correction (BC),  bolometric magnitude ($M_{bol}$), luminosity (L) and classification for MS stars.}
\scriptsize
\begin{tabular}{llccll}
\hline
ID&$\log T_{eff}$&   BC &  $M_{bol}$ & $\log(L/L_{\odot})$&Classification \\
\hline
           4&      3.887&      0.164&      -0.179 &     1.964  & $\delta$ Scuti or New Class \\
           6&      3.833&     -0.113&      -0.286 &     2.007  &$\delta$ Scuti or New Class\\
           7&      3.893&     0.0938&       0.801 &     1.572  & $\delta$ Scuti\\
          13&      4.125&     -0.727&       -2.465 &     2.879  &SPB\\
          14&      3.986&    -0.0234&      -0.509 &     2.097  & New Class\\
          16&      3.966&     0.0234&       0.177 &     1.822  &$\delta$ Scuti or New Class\\
          29&      3.865&      0.133&       0.467 &     1.706  & $\delta$ Scuti\\
          31&      4.196&      -1.273&       -2.258 &     2.796  &SPB\\
          32&      3.893&     0.094&       0.510 &     1.689  &$\delta$ Scuti\\
          34&      3.942&     0.078&       0.142 &     1.836  &$\delta$ Scuti or New Class\\
          47&      3.926&      0.148&       0.351 &     1.752  &$\delta$ Scuti\\
          49&      4.202&      -1.164&       -2.748 &     2.992  &SPB\\
          50&      4.189&      -1.359&        -6.330 &     4.425  &DHcep; Eclipsing binary\\
          54&      4.159&     -0.992&       -3.296 &     3.211  &SPB\\
          55&       4.278&     -1.547&         -4.491&     3.689& $\beta$ Cep \\
          56&       4.045&     -0.414&       -3.660&       3.357& SPB \\
          57&       3.986&     -0.023&      -3.219&      3.181& SPB \\
\hline
\end {tabular}
\end{table}

\section{Variability Characteristics}
The sample phased light curves of identified MS, PMS and field population  are shown
in Figs. 11, 12 and 13 respectively, where
averaged differential magnitude
in 0.04 phase bin along with $\sigma$ error bars has been plotted. The phased light curves of all the MS, PMS and field variable stars are
available online. 

\subsection {MS Variables of the cluster}

Seventeen stars are found to be of MS type stars. The estimated periods of these stars range
between 0.12d to 7.71d. The amplitudes of these variable stars are of a few mag. The location of the
MS variables in the H-R diagram  suggests that 4 stars (13, 31, 49 and 54) could be slow pulsating B (SPB) stars, whereas stars 7, 29, 32 and 47 could be $\delta$ Scuti type stars.
The location of star 14 is found to be between the gap of SPB and $\delta$ Scuti instability strip.
Recently, Mowlavi et al. (2013) have found a large population of new variable stars between SPB stars and the $\delta$ Scuti stars, the region where no pulsations were expected on the basis of theoretical models. 
Four stars namely 4, 6, 16 and 34 lie near the boundary of $\delta$ Scuti instability strip. 
Star 55 could be $\beta$ Cep type variable.

DH Cep (star no 50) = HD 215835 (R.A.=22h 46m 54.11s, Dec = +58$^\circ$ 05$^\prime$ 03.5$^\prime$$\prime$, J2000) is the brightest star in the present sample
and is considered to be the primary exciting star for the H II region S142.
The star is a double-lined spectroscopic and eclipsing binary consisting of two very luminous O5/6 stars (Hilditch et al.  1996) close to the ZAMS. The binary star is an X-ray source, perhaps attributed to colliding winds (Pittard \& Stevens 2002; Bhatt et al.  2010).  
The proper motion of DH Cep is estimated by Baumgardt et al. (2000) as $\mu$ (RA), $\mu$ (DEC) ($-$1.74$\pm$0.84 ,  $-$2.52$\pm$0.81) mas yr$^{-1}$, whereas the SHAO and UCAC3 values are found to be ( $-$4.11$\pm$1.29 ,  $-$2.80$\pm$1.19) mas yr$^{-1}$ and ( $-$2.7 , $-$3.6) mas yr$^{-1}$, respectively.
  The measured radial velocity of DH Cep has been recorded as $-$33$\pm$2 km/s (Hilditch et al.  1996), $-$35.4$\pm$1.8 km/s (Pearce 1949), and $-$39$\pm$3 km/s (Sturm \& Simon 1994; Penny et al.  1997), which is consistent with it being part of the Perseus arm (Georgelin \& Georgelin 1976).
 Our kinematic membership criterion for NGC 7380 region relies on the assumption that DH Cep is a member of the cluster. Our
photometric observations of DH Cep reveals a period of 2.114 day which matches with earlier  period determinations.

 To characterize the variability of stars, in addition to the period, amplitude and shape of light curve of variable stars we need the location of variables in the H-R diagram. Fig. 10 show the H-R diagram for variables along with the theoretical SPB star instability strip, empirical $\delta$ Scuti instability strip and the location of $\beta$ Cep
stars.

Underhill (1969) found that star 56 could be a foreground star of late A to F type, whereas star 57 could be an early
B type star lying at the same 
distance as DH Cep. WEBDA on the basis of the photometric observations gives spectral type of star 57 as B0.5 V.
Though these two stars 56 and 57 are located well away from the  theoretical SPB strip in the H-R diagram, their variability 
characteristics suggest that these could SPB stars.

The spectral types of stars 54 and 55 is estimated by Underhill (1969) as B0.5 and B3 respectively, whereas WEBDA suggests the
spectral types as B1V and B0.5V, respectively. Present photometric observations suggest
the $(B-V)_{0}$ colour of the stars 54 and 55 as -0.164 mag and -0.221 mag which indicate the photometric spectral types as B4.5V and B2.5V.
The $I$ band light curves of stars 54, 55, 56 and 57 suggest that these might be B type stars.

Photometric observations of stars 13, 31 and 49 also suggest that these could be B type MS stars. The spectral type of star 31 reported in WEBDA as B1.5 V.
Their variability characteristics and location in the H-R diagram (cf Fig. 10) suggest that these could be SPB stars. 

The stars 7, 29, 32 and 47 are located in the $\delta$ Scuti instability region (cf Fig. 10). The variability
characteristics of these stars also supports that these could be $\delta$ Scuti stars.  
Star no 14 is located between the gap of SPB and $\delta$ Scuti instability strip in the H-R diagram.
Star numbered 4, 16 and 34 are lying near the boundary of the $\delta$ Scuti instability strip; the periods
of these stars are found to be 4.728 d, 0.406 d and 0.638d, respectively, which are longer than the reported periods of $\delta$ Scuti
stars (0.03d to 0.3 d). The amplitude of these stars are in the range of 0.01 to 0.14 mag. The period and amplitude
of star 14 are 5.462 d and 0.014 mag. 
The proper motions (cf. Table 2) of 4, 14 and 34 suggest that these stars could be the member of cluster NGC 7380. 
We presume that the stars 4, 14, 16 and 34 could belong to new class of variable 
as suggested by Mowlavi et al. (2013). 

\subsection{PMS variables}
The variations in the brightness of both WTTSs and CTTSs  are found to occur at all wavelengths, from X-ray to infrared. The variability time-scale of TTSs ranges from a few minutes to years (Appenzeller \& Mundt 1989). 
The variations in the brightness of TTSs are most probably due to the presence of cool or hot spots on stellar surface and circumstellar disk. The cool spots on the surface of the stars are produced by the emergence of stellar magnetic fields on the photosphere, and are thus indicators of magnetic activity. The cool spots on the photosphere rotate with stars hence are responsible for brightness variation in WTTSs. These WTTSs are found to be fast rotators as they have either thin or no circumstellar disk. The hot spots on the surface of young stars are the consequence of accretion process (Lynden-Bell \& Pringle 1974; Koenigl 1991; Shu et al. 1994). Irregular or non-periodic variations are produced because of changes in the accretion rate. The time-scales of varying brightness range from hours to years. 
The accreting CTTSs show a complex behaviour in their optical and NIR light curves (Scholz et al. 2009).

The present sample consists of 14 probable PMS stars with 4 CTTSs  and 10 WTTSs. The period of these PMS
variables ranges from 0.13d to 53.64 d. The amplitude ranges from 0.008 to 0.371 mag. All of the CTTSs
have longer periods ranging from 5.586 d to 53.644 d, whereas 5 WTTSs (out of 10) have periods less than 2 d.
The amplitudes of CTTSs range from 
0.05 mag to 0.37 mag, whereas most of the WTTSs in the present sample (7 out of the 10) have amplitudes less than 0.12 mag.

The larger amplitude in the case of CTTSs could be due to
presence of hot spots on the stellar surface produced by accretion mechanism. Hot spots cover a small fraction of the stellar surface
but with a high temperature causing larger amplitude of brightness variations (Carpenter et al. 2001). The smaller amplitude in WTTSs suggests
dissipation of their circumstellar disks or these stars might have cool spots on their
surface which are produced due to convection and differential rotation of star and magnetic field.
 This result is
 in agreement with that by Grankin et al. (2007, 2008) and by Lata et al. (2011, 2012). 

\subsection{Field population}
The present sample contains 26 variables in the field. They have periods ranging from 0.05 d to 60.0 d which could belong to the field star population
towards the direction of NGC 7380.
The light curve of star 9 is similar to the RR lyrae type variables.
 RR Lyrae variables pulsate with a period in the range between 0.2 and 1 day. RR Lyrae variables are old, low-mass pulsating stars. 
The characteristics of star 5 (period$\sim$ 0.32d, amp$\sim$ 0.12 mag) and 9 (period $\sim$ 0.26d, amp$\sim$ 0.08 mag)
are consistent with being RR Lyrae variables. 
Stars 21, 22, 35, 38, 39, 42, 43 and 48 have periods greater than 10 days.

%**********************************************************************************************************
\begin{figure}
\includegraphics[width=8cm, height=8cm]{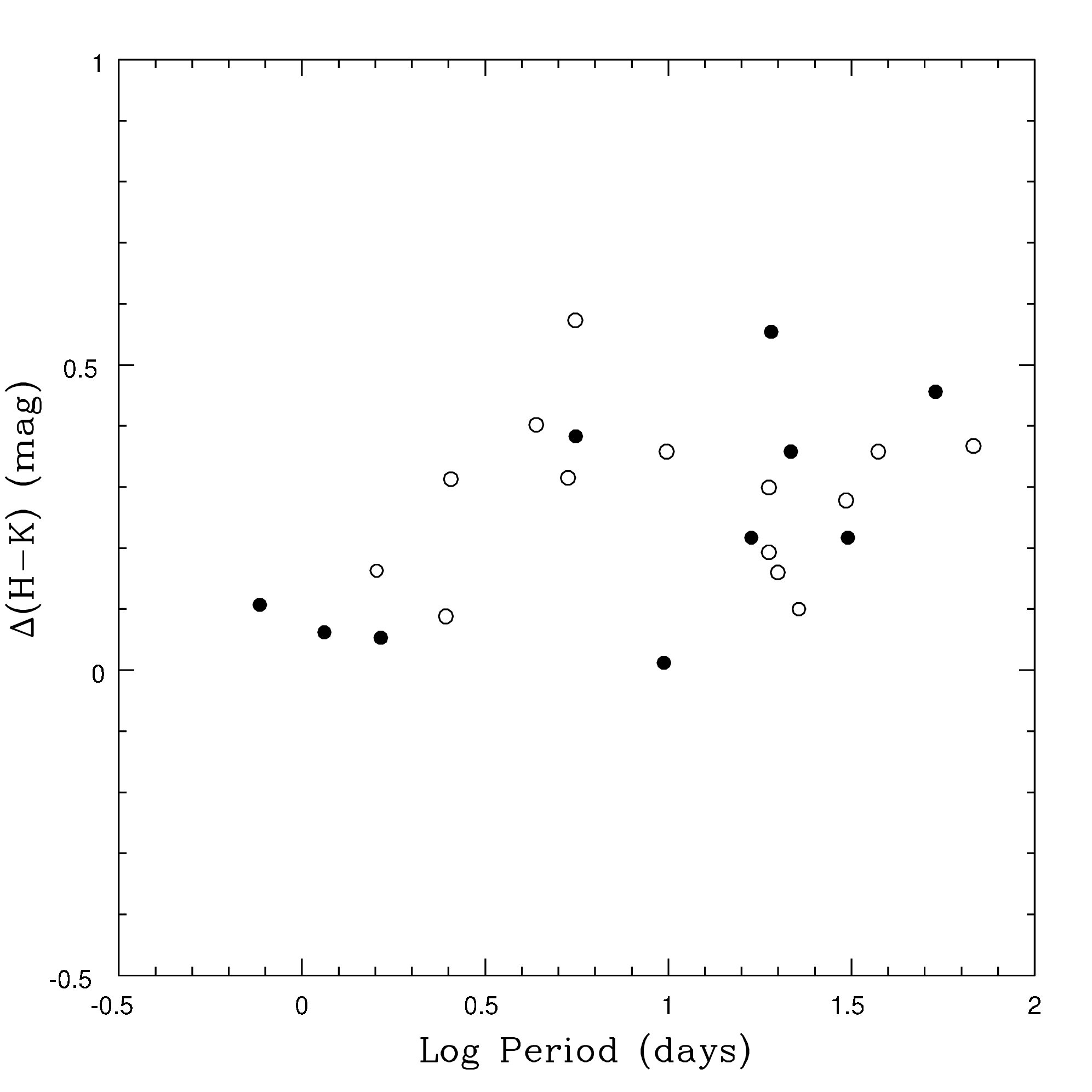}
\caption{Rotation period vs $\Delta$$(H-K)$. The filled and open circles represent data for NGC 7380 and NGC 1893 clusters, respectively.}
\end{figure}
%**********************************************************************************************************

\section {Correlation between rotation and $\Delta$$(H-K)$}
A relation has already been found between the IR excess and the rotation rate for young stars. For example, Rebull et al.  (2014)  
have found  the IR excess does not necessarily imply longer periods,
but a star with a longer period is more likely than those with shorter 
periods to have an IR excess. 
It is suggested that long period stars with little or no IR excess may have just recently cleared their disks 
and have not yet spun up in response to contraction on their way to the ZAMS.
The NIR excess (e.g., $(H-K)$ excess) is a useful indicator for the presence of disk, hence can be used to look for the correlation between  rotation and 
the presence/ absence of accretion disks. 
Here we define a parameter $\Delta$$(H-K)$ as the horizontal displacement from the left reddening vector of `F' region 
in the NIR-CC diagram of identified YSOs as shown in Fig. 6 assuming that $\Delta$$(H-K)$ indicates NIR excess.  Fig. 14  plots  the relationship 
between $\Delta$$(H-K)$ and rotation period of the YSOs.  The data points for NGC 7380 are shown with filled circles whereas open circles 
represents data points for the young cluster  NGC 1893 taken from our earlier 
work (Lata et al. 2014; Pandey et al. 2014). The variables having rotation period $\le$ 0.5 d have not been included in the Fig. 14. There seems to be a correlation between NIR excess and period in the sense that YSOs having higher NIR excess, i.e., having disk, exhibit a longer rotation period. 
There seems to be a sudden change in the $\Delta$$(H-K)$ at $\log$ P $\sim$ 0.4 i.e $\sim$ 2.5 d.

\section {Correlation among rotation, mass, age, and amplitude}
To understand the correlation between mass/age of the TTSs and the rotation period we plot rotation period as a 
function of mass and age in Fig. 15. To increase the sample we have also included data for Be 59 from Lata et al. (2011) 
and NGC 1893 from Lata et al. (2012) because these clusters have similar environments, and moreover the same 
technique has been used to identify variable stars and to determine their physical properties. 
We have considered only those stars which have rotation period $>$ 0.5 d because rotation periods shorter than about 0.5 d would results in rotational velocities
likely exceeding the break-up speed of T Tauri stars.
 Although there is a large scatter, the stars having masses $\gtrsim$ 2 $M_{\odot}$ ($\log M/M_{\odot}$ =0.3) are found to 
be fast rotators. Similarly, the stars having ages $\gtrsim$3 Myr  also seem to be fast rotators. This result is compatible with the disc locking model in which it is expected that stars with disc rotate slowly than those without (Edwards et al. 1993; Herbst et al. 2000; Littlefair et al. 2005).
Fig. 16 plots amplitude  of TTSs variability as a function of  mass/ age of the YSO which, though with large scattering, reveals that amplitude of 
TTSs variability is correlated with the mass (left panel) and age 
(right panel) in the sense that amplitude decreases with increase in 
mass as well as age of a variable star. Deviations  of individual sources 
of Be 59 and NGC 1893 have already been discussed in our earlier works (Lata et al. 2011, 2012).
The decrease in amplitude could be due to the dispersal of the disk. Present  result further supports our previous studies (Be 59: Lata et al. 2011; NGC 1893: Lata et al. 2012) that the disk dispersal mechanism is less efficient for relatively low mass stars. Right panel of Fig. 16 suggests that significant amount of the disks is dispersed by  $\lesssim$5  Myr. This result is in accordance with the result obtained by Haisch et al. (2001).
%*************************************************************************************************************************************************
\begin{figure}
\includegraphics[width=9cm, height=10cm]{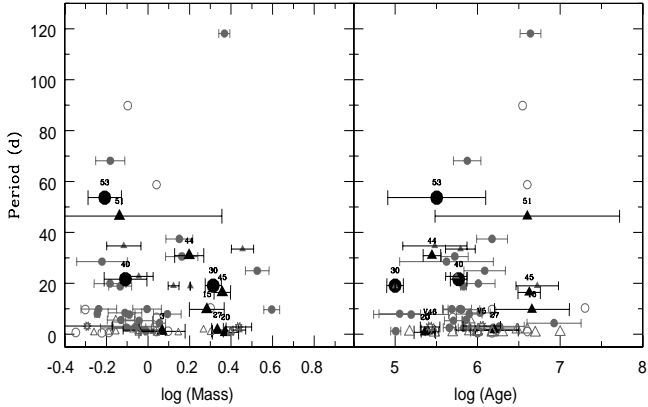}
\caption{Rotation period of TTSs as a function of mass and age. Filled circles (larger; CTTSs) and triangles (larger; WTTSs) represent present data. Open and filled circles (CTTSs), and open, filled triangles (WTTSs) and starred circles (probable YSOs in NGC 1893) represent data for Be 59 and NGC 1893 taken from Lata et. al. (2011, 2012).}
\end{figure}
%*************************************************************************************************************************************************
\begin{figure}
\includegraphics[width=9cm, height=10cm]{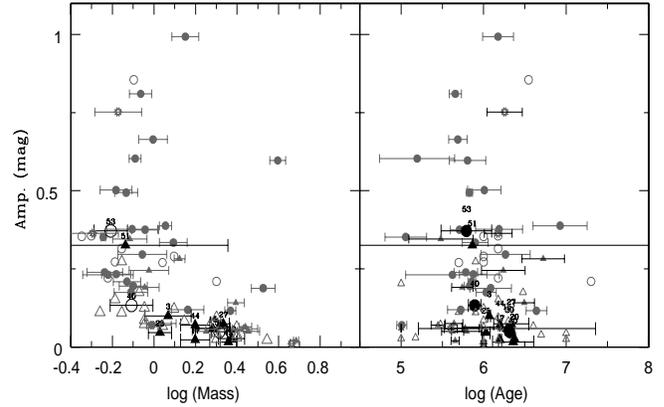}
\caption{Amplitude of TTSs as a function of mass and age. Filled circles (larger; CTTSs) and triangles (larger; WTTSs) represent present data.  Open and filled circles (CTTSs), and open, filled triangles (WTTSs) and starred circles (probable YSOs in NGC 1893) represent data for Be 59 and NGC 1893 taken from Lata et. al. (2011, 2012).
}
\end{figure}

%*************************************************************************************************************************************************

\section{Summary}
This work presents 57 variable stars in the cluster
region of the NGC 7380. 
This study  contains 14 PMS stars. The ages and masses of
the majority of these PMS sources are found to be $\lesssim$ 5 Myr and in the
range 0.60 $\lesssim M/M_{\odot} \lesssim$ 2.30, respectively and hence
these could be T-Tauri stars. Four and 10 PMS stars are classified as CTTSs  and WTTSs respectively. The periods of these PMS
variables range from 0.13 d to 53.64 d. The amplitudes range from 0.008 mag to 0.371 mag.
In addition we have found 17 MS variable stars (SPB stars, $\delta$ scuti, $\beta$ Cep and new class variable stars) and  26 variable stars belonging to the
field star population. These are categorized on the basis of CMDs and TCDs. Variability characterisation has been done 
on the basis of period, amplitude, shape of light curves and location on the H-R diagram.  

\section{Acknowledgment}
The authors are very grateful to the anonymous referee for a critical reading of the paper and useful comments.
This study makes use of data products from the Wide-field Infrared Survey Explorer, which is a joint project of the University of California, Los Angeles, and the Jet Propulsion Laboratory/California Institute of Technology, funded by the National Aeronautics and Space Administration. 

\bibliographystyle{mn2e}
%\bibliography{myref}

\end{document}